\pdfoutput=1
\RequirePackage[2018-12-01]{latexrelease}
\documentclass{rQUF2e}
\renewcommand{\draftnote}{}

\usepackage{epstopdf}
\usepackage{subfigure}
\usepackage{booktabs}
\usepackage{amsmath}
\usepackage{amssymb}
\usepackage{bm}
\usepackage[section]{placeins}
\usepackage{float}
\usepackage{xcolor}

\definecolor{linkblue}{RGB}{0,72,140}
\newcommand{\linkdest}[1]{\pdfdest name{#1} xyz}
\newcommand{\linkref}[1]{\leavevmode\begingroup\color{linkblue}\pdfstartlink attr{/Border [0 0 0]} goto name{#1}\ref{#1}\pdfendlink\endgroup}
\newcommand{\externallink}[2]{\leavevmode\begingroup\color{linkblue}\pdfstartlink
attr{/Border [0 0 0]} user{/Subtype /Link /A << /S /URI /URI (#1) >>}#2\pdfendlink\endgroup}
\makeatletter
\renewcommand{\hyper@natanchorstart}[1]{\pdfdest name{bib:#1} xyz}
\renewcommand{\hyper@natanchorend}{}
\renewcommand{\hyper@natlinkstart}[1]{\begingroup\color{linkblue}\pdfstartlink attr{/Border [0 0 0]} goto name{bib:#1}}
\renewcommand{\hyper@natlinkend}{\pdfendlink\endgroup}
\renewcommand{\hyper@natlinkbreak}[2]{#1}
\makeatother

\begin{document}

\title{Latent Continuum of Regimes in Limit Order Book Dynamics}

\author{
Anjali Thawait\thanks{Email:
\externallink{mailto:anjali.thawait@partner.kit.edu}{anjali.thawait@partner.kit.edu}\newline The author acknowledges
Dr.\ Maxim Ulrich (KIT) for his valuable support and assistance throughout
this research.}\\
\affil{Karlsruhe Institute of Technology (KIT)\\
Karlsruhe, Germany}
}

\maketitle

\begin{abstract}
Market-regime models typically assume a finite set of discrete latent states. We examine whether high
frequency limit-order-book dynamics exhibit distinct regime separation or apparent regimes result from
discretising an underlying continuum of market conditions. We analyse four years of deep limit-order
book data for EURO STOXX 50 index futures across 987 clean trading days from 2022 to 2025, using 30, 45 and 60-second aggregation windows. Market states are represented by \(15 \times 15\) symmetric positive
definite covariance matrices constructed from price, liquidity, depth, book-shape and order-flow variables
and analysed under Log-Euclidean and affine-invariant geometries; the resulting state cloud has an
effective dimension of approximately 1.4. A causal anomaly layer filters statistical, calendar and rollover
contamination, while label-free economic validation evaluates market-state separation independently of
volatility-based proxy labels. A 17-method zoo spanning conventional and Riemannian representations,
reinforced by full-cloud geometric certificates, consistently favours a continuum over a discrete regime
structure. A dominant latent coordinate captures between 83.66\% and 85.27\% of variation in the covariance-state geometry
and tracks VSTOXX without regime labels. This geometric stress axis remains stable across years, although
its average level shifts with market conditions, indicating that conventional regimes are coarse
quantisations of the underlying continuum. A divisive hierarchy spanning macro, macro-subregime, micro
and micro-subregime tiers is evaluated using expanding-year walk-forward tests.
Out-of-sample performance is strongest at the macro tier, while finer tiers fail to generalise. Continuous
fine-scale information remains predictive; its discretisation causes the performance loss. The stress dial is
highly persistent, yet deviations gradually return toward their typical level, halving within approximately
5-15 minutes. Forecasting performance peaks near ten minutes, reaching a maximum pooled
\(R^2_{\mathrm{OOS}}\) of 0.7483. Although incremental predictive value over volatility persistence
strengthens toward one hour, it provides no improvement over HAR at the next-day horizon. Overall, the
latent stress coordinate defines a stable, low-dimensional geometric continuum that captures market
stress and its predictive dynamics. Repository:
\externallink{https://github.com/anjalitwt009/Latent-Continuum-of-Regimes-in-Limit-Order-Book-Dynamics}{github.com/anjalitwt009/Latent-Continuum-of-Regimes-in-Limit-Order-Book-Dynamics}.
\end{abstract}


\begin{classcode}
C22; C38; C55; C58; G17
\end{classcode}

\section{Introduction}\label{sec:introduction}

High-frequency market microstructure is shaped by the continuous interaction between liquidity
supply and demand. Every order submission, cancellation and execution redistributes
liquidity at the best quotes and across deeper price levels. In a deep limit order book (LOB), the
accumulation of these events generates persistent patterns in liquidity, volatility, imbalance,
resiliency and price impact. These recurring market configurations are commonly described as
regimes. Markov-switching and hidden-state models formalize this representation through a finite
collection of latent states, defined as calm, active and stressed or liquid and illiquid,
and probabilistic transitions among them \citep{Hamilton1989,AngTimmermann2012}. However, these regimes
may be coarse quantisations of an underlying geometric continuum rather than naturally distinct
market states. The central empirical question is: does high-frequency LOB
activity form distinct latent states, or are conventional regimes segments of an underlying
continuous state?

The geometry of market state determines how market dynamics are interpreted and forecast. A
discrete representation models market changes as transitions among a finite set of labels with
regime-specific distributions, evaluated through transition probabilities, dwell times and
classification accuracy. A continuous representation models the market as a trajectory on a
low-dimensional latent manifold, examined using latent coordinates, geodesic velocity,
mean-reversion dynamics, relaxation timescales and the horizons over which the state retains
predictive information. The choice of representation determines how we measure persistence,
identify structural changes, define anomalous states and evaluate predictive information. We determine the appropriate
market-state representation by testing whether the LOB state space is discrete and whether its
continuous geometry has predictive value across multiple timescales.
Established microstructure research characterizes LOB variables
as complementary dimensions of liquidity and short-horizon market dynamics: the bid-ask spread
captures the cost of immediacy; displayed depth measures available liquidity; book imbalance and
order-flow pressure reflect directional trading pressure; and price-impact sensitivity captures
the response of prices to trading activity \citep{Kyle1985,Cont2014,Bugaenko2020}. We integrate the
time-varying cross-dependence among these signals into a multivariate geometric representation that
connects observable order-book activity to the latent market state.

We examine the geometry of this multivariate LOB state using four years of nanosecond-stamped,
ten-level market-by-price FESX data, comprising 987 clean trading days from 2022-2025. At 30, 45 and 60-second durations, covariances among 15 microstructure variables form \(15\times15\)
symmetric positive-definite (SPD) market-state representations. Predictive relevance is tested using purged expanding-year
out-of-sample evaluation over horizons from the next native window to one hour, with a separate
next-day HAR benchmark. Each duration is constructed independently
using its own causal, native-grid anomaly mask. These covariance states lie on the SPD manifold,
whose intrinsic geometry differs from ordinary Euclidean space. We analyse this state space using the affine-invariant
(AI) Riemannian metric as the primary geometry and the Log-Euclidean (LE) metric as a transparent diagnostic
\citep{Pennec2006}. This dual-metric design evaluates whether
the empirical conclusions reflect the intrinsic structure of the market-state manifold and remain
robust across alternative geometric specifications.
Before estimating the market-state geometry, we combine matrix-profile discords
\citep{Yeh2016} and Isolation Forest outliers \citep{Liu2008} with deterministic calendar
exclusions to construct duration-specific clean-state masks. Manifold velocity and Hawkes
excitation are retained as measures of rapid market-state
changes and event clustering \citep{Hawkes1971,Bacry2015}; they remain outside the exclusion rule
so that genuine episodes of market stress remain part of the empirical state space.

Our analysis begins by testing the underlying structure of the market-state space. Geometric,
spectral, topological and stability checks determine whether the masked SPD cloud contains
reproducible islands or a connected ridge. A method-zoo comparison then tests whether the inferred
structure remains consistent across clustering, hidden-state, jump, dimensionality-reduction,
nonlinear-manifold and Riemannian representations. After establishing this structure, we interpret
the dominant coordinate using external market indicators. This principal direction accounts for
between \(83.66\%\) and \(85.27\%\) of the variation in market states, while the market-state space
has an effective dimension of approximately \(1.4\). Although constructed without volatility
labels, the axis closely tracks VSTOXX and remains stable across years, while the market's position
along it shifts over time. These findings support interpreting the axis as a continuous stress dial
with a stable orientation that captures changes in market conditions.

Within this continuous framework, we represent discrete regimes as hierarchical level-set bands
along the stress index. These bands provide operational thresholds, such as tighter limits during
high-stress states, without implying fixed market-state boundaries. We evaluate the out-of-sample
robustness of this multilevel structure across four levels of granularity: 4, 8, 16 and 27 bands.
Although finer partitions improve in-sample fit, they transfer poorly out of sample. The continuous
coordinate retains fine-scale information obscured by discretisation, indicating that the weak
out-of-sample transfer results from excessive partitioning rather than the high temporal resolution
of the data.

The continuous stress index has clear temporal dynamics. Half of the deviation from its typical
level disappears within approximately 5-15 minutes. Similar estimates are obtained from the
Ornstein-Uhlenbeck model \citep{Uhlenbeck1930} and an independently estimated transfer operator.
The predictive value of the stress index varies across
forecast horizons. In purged expanding-year out-of-sample tests, pooled \(R^2\) peaks near the
10-minute horizon at approximately 0.68, 0.73 and 0.75 for the three covariance durations. Relative
to a volatility-persistence benchmark, its incremental predictive value increases across intraday
horizons but does not carry over to longer forecast horizons. At the next-day horizon, adding the
stress index does not improve forecasts relative to the heterogeneous autoregressive (HAR) model,
a standard benchmark that combines daily, weekly and monthly measures of realised volatility
\citep{Corsi2009}. Event studies and Granger-causality tests \citep{Granger1969} indicate a strong
in-sample relationship; the index's velocity and acceleration provide no robust out-of-sample
leading signals. Overall, the stress index provides a continuous, economically interpretable
signal for monitoring and forecasting evolving intraday risk.

\subsection{Contribution}

A central methodological contribution is a duration-native multivariate representation of the
limit order book, with covariance states and anomaly masks constructed independently on native
30, 45 and 60-second grids to avoid distortions from resampling a preferred grid. The temporal
and multivariate detectors capture distinct irregularities, while alternative anomaly treatments
do not materially change the economic validation. Exact timestamp alignment and paired
masked-unmasked analyses provide an auditable robustness check.

Second, we evaluate the finite-state regime assumption by examining whether the covariance state
  naturally forms discrete clusters or a continuous geometric structure. Conventional studies
  often select a clustering or hidden-state model first and then interpret the resulting labels as
  evidence that regimes exist. We reverse this sequence by testing the geometry before assigning
  economic labels. A secondary method zoo comprising 17 approaches compares conventional clustering,
  hidden-state and jump models, dimensionality-reduction techniques, nonlinear manifold methods,
  and Riemannian approaches on a common deterministic subsample. The method zoo evaluates
  robustness across alternative representations. We use complementary falsification tests to
  determine whether the market-state space is discrete or continuous.
  These methods test whether variation follows a few dominant directions, forms a connected
  single-peaked structure, contains persistent transition modes, and supports reproducible
  partitions. The results support a low-dimensional continuum.

Third, the paper converts the geometric continuum into an interpretable stress index. The dominant
coordinate is estimated from covariance geometry without regime or volatility labels, with its
sign oriented using training-period VSTOXX. Its stability across years, association with VSTOXX,
and monotonic ordering of future realised volatility provide economic validation and distinguish
a persistent geometric direction from changing stress levels. Within this continuous
interpretation, discrete regimes are represented as hierarchical level-set partitions spanning
macro, macro-subregime, micro, and micro-subregime levels, reaching 27 Log-Euclidean and 31
affine-invariant bands. Although finer partitions improve in-sample separation, their purged
out-of-sample transfer deteriorates as the state distribution migrates. This deterioration reflects
excessive discretisation rather than an absence of high-frequency information.

Fourth, the paper characterises the dynamics and forecast horizon of the continuous state. A
causal slow-fast decomposition shows that changing the smoothing timescale redistributes
predictive information between components, while their combined pooled incremental \(R^2\)
remains close to 0.042 across the evaluated timescales. Ornstein-Uhlenbeck and transfer-operator
estimates independently indicate that deviations from the typical stress level halve within
approximately 5-15 minutes. This agreement indicates that stress shocks dissipate gradually over
short horizons, consistent with mean reversion rather than persistent regime shifts. A panel
comprising Bayesian online change-point detection, CUSUM, EWMA, MMD, and geometry-based control
charts evaluates changes along the continuum.
Forecasting uses a fixed predictor set comprising realised volatility, slow and fast stress
coordinates, and state velocity. Purged expanding-year tests show that the geometric state adds
significant predictive information in every feasible duration, horizon, and metric cell, with the
strongest performance near the 10-minute horizon.

Finally, the paper establishes the empirical boundaries of these results. Longer covariance windows
improve absolute forecast fit but generally reduce the incremental information supplied by the
geometric state. At the next-day horizon, geometric variables provide no robust gain over HAR-LOB
across QLIKE, Mincer-Zarnowitz, Diebold-Mariano, Model Confidence Set and permutation tests.
Event-study and Granger associations do not translate into geometry-robust
out-of-sample velocity or acceleration leads, while critical-slowing-down and Hawkes tests provide no calibrated early-warning or
reflexivity evidence. Robustness checks show that the continuum conclusion survives native-duration
reconstruction, alternative geometries, and alternative anomaly treatments. The resulting
``spread lies'' pattern shows that quoted spread alone can misrepresent liquidity conditions: it
remains narrow even as depth deteriorates and price impact, covariance intensity, and future
volatility signal greater stress. Together, these results identify the geometric state as an intraday stress measure rather
than a next-day forecasting factor, causal signal, or crisis-warning device. Because the evidence
is based on one futures contract over 2022-2025, cross-market universality remains an open
empirical question.

\ifdefined\thesisversion
The remainder of the thesis is organised as follows. Chapter 2 reviews the
related literature and identifies the research gap. Chapters 3--5 describe the
data, covariance-state construction, Riemannian representation and anomaly
gate. Chapters 6--8 develop the continuum, hierarchy, state-dynamics,
forecasting and inferential frameworks. Chapter 9 presents the empirical
results, followed by robustness analysis in Chapter 10 and economic
interpretation in Chapter 11. Chapter 12 concludes. The appendices provide
feature inventories, formula derivations and the complete algorithmic
specification.
\else
The remainder of the paper is organised as follows. Section 2 reviews the related literature and
identifies the research gap. Section 3 describes the FESX data, feature construction, native
observation grids, covariance states and realised-volatility targets. Section 4 develops the
Riemannian state representation and metric methodology. Section 5 presents the unsupervised
anomaly-detection and state-selection framework. Section 6 defines the continuum falsification
tests, method zoo, hierarchical level sets, stress-index dynamics and online monitoring methods.
Section 7 sets out the intraday and next-day forecasting frameworks, hierarchical transfer tests,
leading-signal analysis, chronological validation and statistical inference. Section 8 integrates
the empirical results, validation and robustness evidence across the covariance foundation,
anomaly gate, continuum certificates, hierarchy, continuous dynamics, forecasting and causal
falsification. Section 9 concludes. The Appendix provides supplementary definitions, diagnostics
and robustness results.
\fi
\section{Literature Review}\label{sec:literature}

Limit-order-book research shows that liquidity and price formation extend beyond the best quote.
Order-flow imbalance explains short-horizon price changes, while price-impact measures capture
the response of prices to order flow. We combine these measures in a 15-dimensional feature vector and
represent each market state by its covariance matrix, retaining feature variability, dependence,
and scale changes associated with stress. This is particularly relevant in tick-constrained
markets, where the quoted spread may remain fixed while depth, imbalance, and price impact
deteriorate. Unlike deep LOB models that map tensors directly to returns or price direction
\citep{Zhang2019DeepLOB,briola2024deeplob}, we treat forecasting as a separate downstream test.
Oracle Approximating Shrinkage stabilises covariance estimates from short windows by shrinking
them toward a scaled identity matrix \citep{Chen2010}, producing reliable SPD states for geometric
analysis.

Covariance matrices lie on the curved space of symmetric positive-definite matrices, where
Euclidean operations can produce invalid paths or averages. Riemannian methods instead define valid
distances, means, and statistical coordinates directly on this space. Log-Euclidean geometry maps matrices into efficient
log coordinates that preserve positive definiteness and make distance exactly Euclidean
\citep{Arsigny2006}. Affine-invariant geometry compares covariance matrices after whitening and is
robust to linear rescaling and recombination. We therefore specify scaling, standardisation,
reference points, and metric choice explicitly, using AI as the primary geometry and LE as the
diagnostic comparison.

Anomaly detection in high-frequency financial data distinguishes data irregularities from genuine
extreme states. Temporal detectors identify atypical trajectories, while multivariate detectors
identify rare feature configurations, with performance varying across datasets and evaluation
criteria \citep{rewicki2023anomaly,qiu2025tab}. Matrix-profile discords detect unusual temporal
subsequences, whereas Isolation Forest identifies unusual multivariate combinations. Although
financial applications often target manipulation, fraud, or
labelled abnormalities \citep{Bao2025,Lin2025,Letteri2025}, unusual observations in geometric state
analysis may reflect genuine stress. State selection therefore combines complementary detectors
with deterministic calendar information.

Finite-state models estimate latent regimes and transitions for forecasting and execution.
Hierarchical HMMs introduce nested substates
\citep{wisebourt2011hhmm}, jump models identify structured patterns \citep{aydinhan2024jump}, and
regime-aware environments condition trading policies \citep{Amrouni2022}; however, these methods
impose a finite-state structure before testing partition stability. Continuous-state methods
instead recover low-dimensional coordinates and metastable dynamics without fixing a regime count
\citep{CoifmanLafon2006,Roeblitz2013}, allowing discrete regimes to be interpreted as level-set
bands of a continuum. Our framework tests discreteness first, treats internal clustering criteria
as descriptive, and evaluates hierarchical granularity through chronological out-of-sample
transfer. Slow and fast continuous components are examined for predictive information obscured by
increasingly fine partitions.

Clustering algorithms produce a partition whenever a regime count is imposed, even when the data
form a connected continuum, so cluster labels alone cannot establish discrete market states. We
therefore test properties expected under genuine separation: effective dimension measures dominant
directions, unimodality tests for distinct peaks \citep{HartiganDip1985}, \(H_0\) topology
identifies disconnected components, and \(H_1\) topology detects material loops
\citep{Edelsbrunner2002}. Transfer-operator spectra assess persistent transition modes, while
day-block bootstrap stability tests whether partitions remain reproducible across resampled
trading days \citep{Hubert1985}. The conclusion combines these complementary
diagnostics rather than relying on one statistic. A method zoo then compares conventional
clustering, hidden-state and jump models, dimensionality-reduction methods, nonlinear manifolds,
and Riemannian representations on a common deterministic sample. It tests representation
robustness without selecting a preferred clustering algorithm or replacing the full-cloud
geometric tests.

Establishing that market states form a continuum does not reveal how stress evolves over time.
The Ornstein-Uhlenbeck process tests whether the stress dial returns toward a typical level and
estimates how quickly temporary shocks decay through the relaxation time and half-life. These
estimates connect stress persistence to the relevant forecasting horizons; they are not treated
as a standalone test for structural regime shifts. Transfer operators
provide a nonparametric comparison, with leading eigenvalues identifying slow modes, metastability,
and relaxation timescales; agreement between the two approaches strengthens the evidence for robust
dynamics. A causal slow-fast decomposition separates persistent state variation from short-lived
deviations and tests predictive information across smoothing timescales. Online monitoring combines
BOCPD \citep{AdamsMacKay2007}, CUSUM \citep{Page1954}, EWMA \citep{Roberts1959}, and Riemannian
distance from a rolling LE or AI Fr\'echet mean. These methods identify movement along and around
the continuum without changing the anomaly definition or regime representation, while causal
out-of-sample comparisons distinguish persistent dynamics from statistically detectable structural
shifts.

Forecast evaluation must account for horizon-specific predictive value: intraday forecasts are
assessed using overall accuracy and improvement over persistent realised volatility
\citep{Andersen2003,BarndorffNielsen2002}, while next-day forecasts measure incremental performance
over HAR. Comparisons use robust losses, predictive-accuracy tests, and
multiple-model controls \citep{Patton2011,DieboldMariano1995,Hansen2011}. Leading-signal claims
combine event studies, Granger tests, and out-of-sample tests of stress level,
velocity, and acceleration. Critical-slowing-down indicators are calibrated against random
non-event periods. We therefore evaluate stress measurement, intraday prediction, next-day
improvement, and transition signals through dedicated empirical tests.

Recent work applies geometry and representation learning to financial regimes, but most studies
begin by assuming that distinct states exist and then estimate their number, boundaries, or
transitions. Hidden-Markov, jump, clustering, and regime-aware learning models all produce state
labels once a finite structure is imposed. SPD and Riemannian methods similarly use covariance
matrices for regime detection or manifold clustering \citep{orton2024spdnet,shi2025frechet}, while
geometric observables have been proposed as regime indicators \citep{hammond2026geometric}. Neural
hidden-Markov and architecture-search methods extend this approach to high-frequency order flow
\citep{hu2026neuralhmm,devadiga2025regimenas}. These methods describe assigned regimes, but leave a
more basic question unresolved: does the underlying market-state space contain genuine boundaries,
or do the labels divide a connected continuum?

Our research addresses this gap using four years of FESX data. Applications of SPD geometry to
financial regimes remain limited, with no existing approach combining a duration-native LOB
covariance state with a direct geometric test of discreteness. We integrate duration-native
covariance states and anomaly masks with an
AI-primary/LE-diagnostic Riemannian design, a set of predeclared falsification tests, and a
geometry-derived stress coordinate constructed independently of regime and volatility labels. The
contribution lies in applying Riemannian covariance geometry to high-frequency LOB states instead of
introducing another clustering or direct-prediction model. Hierarchical bands are constructed after
the continuum test and evaluated through chronological out-of-sample transfer. Continuous-state
dynamics are then examined through slow-fast decomposition, mean reversion, transfer timescales,
and online change detection. The results support a connected and effectively low-dimensional
market-state continuum under both geometries. Overall, this study positions Riemannian LOB geometry as a framework
for testing market-state structure while clearly separating state measurement from predictive and
causal claims. The contribution is therefore a test-first framework for determining the structure
of high-frequency market states and for separating what the covariance geometry measures from what
the evidence supports about forecasting and causality.

\section{Market, Data and State Construction}\label{sec:data}

We study the EURO STOXX 50 index future (FESX) traded on Eurex from 2022 to 2025, using the
liquid front contract over 16 quarterly expiries. The raw feed contains nanosecond-stamped,
ten-level order-book observations of bid and ask prices, quantities, and trade volume,
which we aggregate into fixed-duration windows for analysis. We supplement these data with daily
VSTOXX closing levels, matched by trading date and used as an external market-stress measure for
economic validation and event-based analysis rather than covariance-state construction. The initial
\(1{,}023\) exchange dates yield \(1{,}018\) valid feature days, of which 987 remain in the final
analytical sample after anomaly filtering and calendar exclusions for opening and closing
half-hours, European Central Bank policy windows, and contract-expiry periods.

\begin{figure}[!htbp]
\centering
\includegraphics[width=0.98\linewidth]{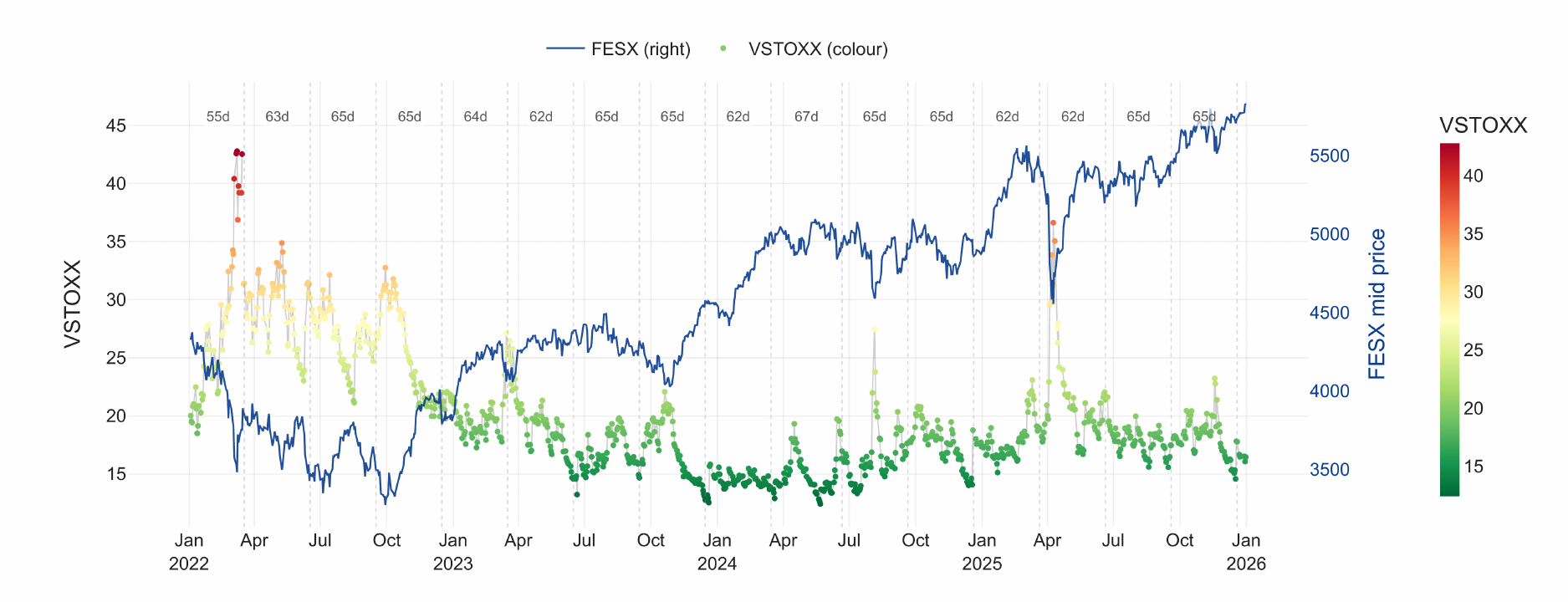}
\caption{FESX contract calendar and daily VSTOXX, 2022-2025. Dotted lines mark quarterly
expiries, and labels report trading days per contract.}
\label{fig:data_contract_calendar}
\end{figure}

Covariances, features, and anomaly indicators are constructed separately for each duration. A
complete session contains \(1{,}021\) windows at 30 seconds, \(681\) at 45 seconds, and \(511\) at
60 seconds. After anomaly filtering, the samples contain \(837{,}890\) states at 30 seconds,
\(559{,}261\) at 45 seconds, and \(419{,}493\) at 60 seconds, retaining \(80.71\%\), \(80.76\%\),
and \(80.81\%\) of valid states, respectively.

For duration \(h\in\{30,45,60\}\) seconds, the unmasked state count is
\begin{equation}
N_h^{\mathrm{unmasked}}
=\sum_{d=1}^{1018}\sum_{t\in\mathcal G_{d,h}}
\mathbf 1\!\left(n_{d,t}\geq17\right)
\label{eq:data_unmasked_count}
\end{equation}
where \(\mathcal G_{d,h}\) is the set of windows on day \(d\) at duration \(h\), and \(n_{d,t}\) is
the number of valid order-book observations in window \(t\). With 15 variables, the threshold of 17
provides one observation beyond the minimum of 16 required for a full-rank sample covariance.

The window store contains 143 variables covering prices, spreads, ten-level depth,
order-flow imbalance, trading activity, realised variation, jump measures, liquidity,
calendar position, lags, and robust rolling statistics. Level variables retain their
history, while change-based variables restart each day so overnight movements are excluded
from intraday analysis. Positive heavy-tailed auxiliary variables use \(\log(1+x)\), while
signed heavy-tailed variables use \(\operatorname{asinh}(x)\). Rolling standardisation uses
the median and median absolute deviation:
\begin{equation}
z_t=\frac{x_t-\operatorname{med}_{R}(x)}
{1.4826\,\operatorname{MAD}_{R}(x)+10^{-9}}
\label{eq:data_robust_z}
\end{equation}
where \(R\) is the rolling historical reference period used to calculate the median and MAD.

Feature construction uses a continuous, gap-aware, and jump-aware second pass. Level features
remain continuous across sessions, while return, lag, and velocity calculations reset at
session boundaries and detected gaps to prevent overnight changes from entering intraday measures.
Jump awareness separates realised variation into bipower continuous variation and a jump component,
with Lee-Mykland flags and jump ratios identifying discontinuities.
Median-and-MAD rolling baselines prevent isolated jumps from inflating their own reference scale. These jump variables
support anomaly detection and validation but remain outside the 15 inputs used to construct
the covariance state.

\label{subsec:data_covariates}
Each covariance state is constructed from a fixed tick vector \(x_i\in\mathbb R^{15}\), where
\(b_\ell\) and \(a_\ell\) denote bid and ask prices, \(q_\ell^b\) and \(q_\ell^a\) their
quantities at level \(\ell\). The vector contains the best bid and ask,
mid-price, quoted and relative spreads, microprice, level-one bid and ask quantities, trade volume,
five-level depth imbalance, level-one through level-three imbalance, depth-weighted five-level
order-flow imbalance, and clipped price pressure. It combines price level, transaction cost, size,
and directional pressure while retaining a sufficiently low dimension for
short-window covariance estimation. The five-level order-flow imbalance uses a depth-decay
structure.

For native window \(t\), all finite tick vectors are collected in
\(\mathcal X_t=\{x_i:\tau_i\in[t,t+h)\}\). A state is created only when
\(n_t=|\mathcal X_t|\ge d+2=17\). For each eligible window, \(S_t\) denotes
the sample covariance around the within-window mean vector \(\bar x_t\),
\(\mu_t\) denotes the average variance of the 15 features, and \(\alpha_t\)
is the average squared magnitude of all variance and covariance entries in
\(S_t\). Equivalently, \(\alpha_t=\|S_t\|_F^2/d^2\), so it measures the overall
covariance energy used to determine the shrinkage strength. Oracle Approximating
Shrinkage combines \(S_t\) with the scaled identity target \(\mu_t I_{15}\).
The data-dependent shrinkage weight is
\begin{align}
\delta_t&=\min\!\left\{
\frac{\alpha_t+\mu_t^2}
{(n_t+1)(\alpha_t-\mu_t^2/d)},\,1\right\}\label{eq:data_oas_weight}\\
\Sigma_t&=(1-\delta_t)S_t+\delta_t\mu_t I_{15}+10^{-8}I_{15}
\label{eq:data_oas_sigma}
\end{align}
When the denominator in \eqref{eq:data_oas_weight} is non-positive,
\(\delta_t=1\). Shrinkage reduces small-sample eigenvalue dispersion, while the diagonal
floor ensures strict positive definiteness. Verification finds no non-positive eigenvalues across the unmasked
covariance sample, and the implementation agrees with an independent OAS reference to
approximately \(10^{-13}\). Each retained \(\Sigma_t\) is therefore a valid
\(15\times15\) SPD market state with a 120-dimensional symmetric tangent representation.

Within-session log mid-price returns provide the realised-volatility outcomes, with the first
return of each session set to zero to exclude overnight price changes. Realised volatility,
bipower variation, and the non-negative jump component are computed
separately for forecasting and validation and do not enter the 15-variable covariance state
\(\Sigma_t\). Forward volatility is evaluated from the next complete native
window to one hour, using horizons of 30, 45 and 60 seconds and 2, 5, 10, 15, 30 and 60 minutes.
Appendix~\linkref{app:forecast_targets} summarises the realised-variation definitions,
forecast horizons, and alignment constraints.

\section{Riemannian State Representation}\label{sec:method}\linkdest{sec:method}

In our research, we represent each eligible native window with an OAS-stabilised
\(15\times15\) symmetric positive-definite (SPD) covariance matrix, using a fixed feature
order across dates and durations. This matrix serves as the common market-state
representation and retains both individual feature variation and cross-feature dependence.
Ordinary Euclidean comparisons treat matrix entries as unrestricted values and can be
dominated by differences in feature units or price levels. We therefore use Riemannian
geometry, which accounts for the SPD structure and measures relative changes in covariance
eigenstructure. Riemannian distances and means compare market states over time, while tangent
coordinates identify the dominant directions used in the continuum, hierarchy, and
forecasting analyses.

We then apply the matrix logarithm to every OAS covariance state. This transformation converts
the positive eigenvalues to a logarithmic scale and places the covariance states in a flat
coordinate space. For two native windows \(t\) and \(u\), the
Log-Euclidean distance between their OAS covariance states is
\begin{equation}
d_{\mathrm{LE}}(\Sigma_t,\Sigma_u)
=\|\log\Sigma_t-\log\Sigma_u\|_F
\label{eq:method_dle}
\end{equation}
A small distance indicates similar market conditions, while a large distance indicates
substantial changes in the joint behaviour of prices, liquidity, quantities, imbalance, and
order flow. We use this distance to compare individual states, assign states to Log-Euclidean
centres, and measure changes between monthly and yearly market conditions. For a selected
group or period, \(n\) denotes the number of covariance states and \(\Sigma_i\) denotes its
\(i\)-th state. The Log-Euclidean mean is
\begin{equation}
\bar\Sigma_{\mathrm{LE}}
=\exp\!\left(n^{-1}\sum_{i=1}^n\log\Sigma_i\right)
\label{eq:method_le_mean}
\end{equation}
The resulting Log-Euclidean mean summarises each group or period and provides the initial estimate
for the iterative affine-invariant Fr\'echet mean. These quantities establish the geometric
structure underlying the subsequent representation, drift, and forecasting analyses.

In our research, we apply an exact Log-Euclidean tangent isometry by converting each
log-covariance matrix into a compact vector containing its upper-triangular entries:
\begin{equation}
\phi(\Sigma)=\operatorname{vech}_{\sqrt 2}(\log\Sigma)\in\mathbb R^{120},
\qquad
\|\phi(\Sigma_t)-\phi(\Sigma_u)\|_2
=d_{\mathrm{LE}}(\Sigma_t,\Sigma_u)
\label{eq:method_vech}
\end{equation}
The 120 coordinates consist of 15 diagonal and 105 unique off-diagonal entries. Each
off-diagonal element is multiplied by \(\sqrt2\) to preserve the Frobenius norm. The forward
transformation and exact reconstruction are
\begin{equation}
X_t=\phi(\Sigma_t)\in\mathbb R^{120},
\qquad
\Sigma_t=\exp\!\left(\operatorname{unvech}_{\sqrt2}(X_t)\right)
\label{eq:method_le_reconstruction}
\end{equation}
This reversible mapping enables efficient vector computation while preserving exact
Log-Euclidean distances. Euclidean PCA and \(k\)-means on the unscaled vectors are therefore
exact Log-Euclidean operations rather than heuristic vectorisations.

For the affine-invariant analysis, we reconstruct the covariance matrices and compare native
windows \(t\) and \(u\) using
\begin{equation}
d_{\mathrm{AI}}(\Sigma_t,\Sigma_u)
=
\left\|
\log\!\left(\Sigma_t^{-1/2}\Sigma_u\Sigma_t^{-1/2}\right)
\right\|_F
=
\left(\sum_{j=1}^{15}\log^2\gamma_{j,t,u}\right)^{1/2}
\label{eq:method_dai}
\end{equation}
Here, \(\gamma_{j,t,u}\) are the generalised eigenvalues of
\((\Sigma_u,\Sigma_t)\). The whitening transformation expresses
the covariance structure in window \(u\) relative to window \(t\). A small distance
indicates similar relative covariance structures, whereas a large distance indicates
substantial changes in their variance, dependence, or eigenstructure. In our research, this
distance is used to assign states to affine-invariant centres, compare annual market conditions,
measure geometric drift, and assess agreement with the Log-Euclidean results.

For a collection of \(n\) covariance states, the affine-invariant Fr\'echet, or Karcher, mean
is the representative SPD matrix that minimises their total squared affine-invariant distance:
\begin{equation}
\bar{\Sigma}_{\mathrm{AI}}
=
\underset{M\in\mathcal S_{++}^{15}}{\arg\min}
\sum_{i=1}^{n}d_{\mathrm{AI}}^2(M,\Sigma_i)
\label{eq:method_karcher_objective}
\end{equation}
We initialise the calculation with the Log-Euclidean mean and update the estimate iteratively:
\begin{equation}
M^{(r+1)}
=
M^{(r)1/2}
\exp\!\left[
\frac{1}{n}\sum_{i=1}^{n}
\log\!\left(M^{(r)-1/2}\Sigma_iM^{(r)-1/2}\right)
\right]
M^{(r)1/2}
\label{eq:method_karcher_update}
\end{equation}
At each iteration, the covariance states are expressed relative to the current mean, their
tangent displacements are averaged, and the result is mapped back to the SPD space. We use
the resulting Karcher means as affine-invariant cluster centres and as representative states
for period-level drift comparisons.

Repeated OAS-floor eigenvalues destabilise the 45 and 60-second affine-invariant eigen
decompositions. We therefore add a deterministic diagonal perturbation of \(10^{-6}\) to
\(3\times10^{-6}\) times each matrix's mean diagonal. The 30-second matrices converge without
adjustment, and Log-Euclidean coordinates are unaffected. Affine-invariant geometry then tests
robustness to feature rescaling and recombination. For each duration, a global Karcher mean
estimated from up to 40,000 masked states defines the tangent reference point. The Riemannian
logarithm expresses each state \(\Sigma_t\)
as a displacement from this reference:
\begin{equation}
\operatorname{Log}_{M}(\Sigma_t)
=
M^{1/2}
\log\!\left(
M^{-1/2}\Sigma_tM^{-1/2}
\right)
M^{1/2}
\label{eq:method_ai_logmap}
\end{equation}
In the implementation, we use the corresponding whitened tangent coordinate
\begin{equation}
Z_t^{(\mathrm{AI})}
=
\operatorname{vech}_{\sqrt2}\!\left[
\log\!\left(
M^{-1/2}\Sigma_tM^{-1/2}
\right)
\right]
\in\mathbb R^{120}
\label{eq:method_ai_tangent}
\end{equation}
Principal geodesic analysis (PGA), the manifold analogue of PCA, identifies the dominant
direction of covariance-state variation in the Karcher-mean tangent space. Its standardised
AI coordinate is
\begin{equation}
s_t^{(\mathrm{AI})}
=
\left\langle
\operatorname{std}\!\left(Z_t^{(\mathrm{AI})}\right),
u_1^{(\mathrm{AI})}
\right\rangle
\label{eq:method_ai_score}
\end{equation}
where \(u_1^{(\mathrm{AI})}\) is the leading principal geodesic direction.

The operational continuum uses the LE coordinate, whereas AI analysis uses a separately
estimated Karcher-centred PGA coordinate; the two are not interchangeable.
For each duration, all metric-sensitive procedures are applied twice to its native masked cloud.
The AI path reconstructs \(\Sigma_t\), assigns states using \(d_{\mathrm{AI}}\), and updates
Karcher centroids, while the LE path operates directly on the unscaled \(\phi(\Sigma_t)\). A
finding is classified as metric-robust only when both analyses satisfy the same predeclared
criterion. Any disagreement is reported explicitly, while headline inference follows
affine-invariant geometry.
Principal-direction signs are oriented using training-period VSTOXX, which determines only
the sign and does not enter covariance construction. Forecasts use expanding-year folds
with a one-day embargo and training-only estimation; the AI PGA coordinate is re-estimated
within each training fold (standardisation, principal-geodesic axis, and orientation), with
no future observations entering the representation.

\section{Unsupervised Anomaly Detection and State Selection}

Reliable inference on covariance-state geometry requires separating atypical or
non-representative observations from genuine changes in market conditions. We therefore
employ a non-destructive, unsupervised anomaly-detection framework that preserves all
detector scores and flags while defining a reproducible clean-state subset for downstream
analysis. The framework distinguishes statistical outliers and predetermined non-continuous
trading periods from economically meaningful variation in volatility, liquidity, order flow,
trading intensity, and covariance structure.

A statistical anomaly is a window with an unusual trajectory or combination of market
features, while calendar exclusions capture auctions, announcements, contract rollovers,
and incomplete data. The clean-state rule excludes matrix-profile discords, Isolation Forest
outliers, and calendar events that may distort the estimated geometry. Economically
meaningful changes remain part of the analysis. Manifold velocity measures rapid transitions
between consecutive covariance states, while Hawkes excitation captures self-reinforcing
clustering in trading activity. These measures are retained as diagnostic signals because
they describe market dynamics rather than data errors. Since exchange-verified anomaly
labels are unavailable, the framework provides unsupervised state selection rather than
supervised classification of errors, manipulation, or crises.

\subsection{Six detector roles}\label{subsec:method_detectors}

The pipeline computes six components with different statistical purposes and state-selection
roles. This separation prevents valid market transitions and self-exciting trading activity from
being misclassified as data corruption.
Table~\ref{tab:detector_roles} summarises the research use and state-selection role of each
component.

\begin{table}
\tbl{Anomaly-detection components and their roles in state selection.}{
\begin{tabular}{p{3.2cm}p{8.5cm}p{2.0cm}}
\toprule
Component & Research use & Status\\
\midrule
M1: Tick-validity guard
& Prevents invalid ticks from contaminating the one-second series used by MDI.
& Pre-filter\\

M2: Manifold velocity
& Measures rapid transitions in covariance structure.
& Diagnostic\\

M3: Matrix-profile discord
& Detects unusual intraday trajectories that differ from recurring market patterns.
& Gates\\

M4: Hawkes excitation
& Measures self-exciting trading activity and distinguishes endogenous clustering from
exogenous shocks.
& Diagnostic\\

M5: Isolation Forest
& Detects unusual multivariate combinations of liquidity, volatility, depth, flow, and
activity.
& Gates\\

M6: Calendar rule
& Removes intervals not representative of continuous-book trading.
& Gates\\
\bottomrule
\end{tabular}}
\label{tab:detector_roles}
\end{table}

\noindent\textbf{M1: tick-validity guard (pre-filter).}
Before trajectory detection, M1 removes a raw tick if
\[
b_1\le0\ \lor\ a_1\le0\ \lor\ s<0\ \lor\ s>20\
\lor\ |\Delta m|>50
\]
Here, \(b_1\) and \(a_1\) are the best bid and ask, \(s\) is the quoted spread in index
points, and \(\Delta m\) is the change in mid-price. These broad integrity thresholds remove
invalid or extreme ticks before one-second trajectory construction.

\noindent\textbf{M2: Manifold velocity (diagnostic).}
Consecutive native covariance states receive
\begin{equation}
v_t^{(\Sigma)}
=
d_{\mathrm{AI}}(\Sigma_{t-1},\Sigma_t)
\label{eq:method_anomaly_velocity}
\end{equation}
where \(\Sigma_{t-1}\) and \(\Sigma_t\) are the covariance states of two adjacent native
windows, \(t\) indexes time, and \(d_{\mathrm{AI}}\) is the affine-invariant distance
defined in Section~\linkref{sec:method}. Thus, \(v_t^{(\Sigma)}\) measures how far the market's
covariance structure moves from one window to the next. We standardise each affine-invariant
distance against a trailing 120-window baseline, corresponding to 60, 90 and 120 minutes on
the 30, 45 and 60-second grids. Values above \(z_t=3\) are recorded as transition alerts; the
first state of each session is undefined. These alerts remain diagnostic because rapid movement
may represent a genuine market transition rather than a data error. The 30-second Log-Euclidean
velocity is retained as a metric-robustness check.

\noindent\textbf{M3: Matrix-profile detection (gate).}
M3 detects unusual temporal paths rather than isolated feature vectors. After M1 filtering,
raw updates are aggregated to one-second series consisting of median spread, summed trade
count, and mean level-one imbalance. For two \(m\)-second subsequences \(T_i\) and \(T_j\),
their z-normalised distance is
\begin{equation}
D_{ij}=\sqrt{2m(1-\rho_{ij})}
\end{equation}
where \(\rho_{ij}\) is their correlation. The matrix-profile score is
\begin{equation}
P_i=
\min_{\lvert i-j\rvert>\lceil m/4\rceil}D_{ij},
\qquad m=30
\label{eq:method_matrix_profile}
\end{equation}
With \(m=30\), the exclusion zone removes neighbouring matches within eight seconds.
Separate profiles for spread, activity, and imbalance are combined by retaining the largest
value at each second, and the native-window MDI score is their within-window maximum. The
subsequence length is fixed across grids, while aggregation follows the 30, 45 or 60-second
duration. Windows above the daily 95th percentile of positive scores are flagged, identifying
approximately the most unusual 5\% of paths.

\noindent\textbf{M4: Hawkes reflexivity (diagnostic).}
We use an exponential Hawkes process to measure clustering in trade arrivals within
non-overlapping five-minute blocks:
\begin{equation}
\lambda(u)=\mu+\sum_{t_i<u}\alpha e^{-\beta(u-t_i)},\qquad
\eta=\alpha/\beta<1
\label{eq:method_hawkes}
\end{equation}
where \(u\) is elapsed time within the block, \(t_i\) is the timestamp of an earlier trade
event, and \(\lambda(u)\) is the conditional trade-arrival rate. The parameter \(\mu\)
represents baseline trading intensity, \(\alpha\) measures the immediate increase in
intensity following a trade, and \(\beta\) controls how rapidly this effect decays. The
branching ratio \(\eta\) summarises reflexivity. Values near zero indicate predominantly
externally driven arrivals, while values approaching one indicate stronger self-exciting
activity.

We estimate \(\mu\), \(\alpha\), and \(\beta\) by constrained maximum likelihood for
blocks containing at least ten events. The calculation uses decay-rate starting values of
\(0.5\), \(2\), and \(10\), and caps very active blocks at 1,500 events. For block \(w\),
the baseline-count residual is
\begin{equation}
R_w=N_w-\widehat{\mu}_wT_w,
\qquad
z_w^{(R)}=\frac{R_w}{s_R}
\end{equation}
where \(N_w\) is the observed event count, \(T_w\) is the block length, and \(s_R\) is the
within-day residual standard deviation. Blocks satisfying
\(\lvert z_w^{(R)}\rvert>2\) receive an exogenous-shock diagnostic. We retain buy- and
sell-side branching ratios, cross-excitation, and excitation asymmetry to assess whether
activity is baseline-driven or reinforced by recent trades. These diagnostics do not enter
the clean-state gate because self-excitation may represent genuine market behaviour. Results
are interpreted cautiously because the snapshot feed may omit events needed for complete
identification.

\noindent\textbf{M5: Isolation Forest detection (gate).}
Isolation Forest uses 40 engineered window-level features that are separate from the 15
raw inputs used to construct each covariance state. The complete anomaly-feature inventory
is reported in Appendix~\linkref{app:iforest_features}; its jump-aware inputs are the
non-negative jump component and jump ratio.
M5 identifies native windows containing unusual combinations of volatility, liquidity,
impact, order flow, depth, activity, arrival, resilience, and ten-level book-shape
features. For one window \(x\), its conventional isolation score is
\begin{equation}
a(x)=2^{-E[h(x)]/c(N)},
\qquad
c(N)=2H_{N-1}-\frac{2(N-1)}{N}
\label{eq:method_iforest}
\end{equation}
where \(x\) is its standardised 40-variable feature vector, \(h(x)\) is the number of
random tree splits required to isolate it, \(E[h(x)]\) is the average isolation depth
across the forest, \(N\) is the number of complete windows available for that daily fit,
and \(H_{N-1}\) is the corresponding harmonic number. The normalising term \(c(N)\) makes
isolation depths comparable across different sample sizes. Unusual windows
are isolated with fewer splits and consequently receive higher anomaly scores.

For each day and duration, we standardise 40 window-level variables and fit 200 isolation
trees using 1\% contamination and seed 42. The stored negative decision score increases with
anomaly severity; the most unusual 1\% of eligible rows are flagged, while incomplete rows
are excluded without imputation. Isolation Forest detects multivariate configurations
complementary to MDI's temporal paths, as reflected by their low overlap. Its flags enter
the clean-state gate but do not define regimes or forecast volatility.

\noindent\textbf{M6: Calendar and rollover exclusions (gate).}
M6 applies deterministic exclusions to each duration-specific grid. Trading days span
07:00-15:30 UTC, yielding 1,021, 681, and 511 potential timestamps at 30, 45, and
60 seconds. We exclude windows outside 07:30-15:00 UTC, those within 30 minutes of
11:45 UTC on scheduled ECB announcement days, the quarterly FESX expiry date and
preceding trading day, and five truncated or defective dates. These rules remove periods
shaped by auctions, announcements, contract rollover, or incomplete data rather than
continuous-book dynamics.

For each duration, the clean-state mask retains only windows unflagged by the matrix
profile, Isolation Forest, or calendar exclusions; invalid ticks are removed beforehand,
while manifold velocity and Hawkes excitation remain diagnostic. Each mask is constructed
independently on its native grid and applied to the covariance states and aligned arrays
through exact timestamp matching.

Since exchange-verified anomaly labels are unavailable, no detector flag is treated as
ground truth. For each duration, the combined anomaly score is the equally weighted
average of the within-duration percentile ranks of the matrix-profile and Isolation Forest
scores, placing both on a common \([0,1]\) scale. We evaluate separation, overlap, temporal
stability, and economic ordering; own-flag AUROC is used only as a threshold diagnostic.

\subsection{Proxy-label and external economic evaluation}
\label{subsec:method_proxy_evaluation}

As a secondary economic benchmark, we define a proxy event for window \(t\) as
\begin{equation}
Y_t^{\mathrm{proxy}}
=
\mathbf 1\!\left[
\mathrm{RV}_t>Q_{0.99}(\mathrm{RV})
\;\lor\;
\Delta\mathrm{VSTOXX}_{d(t)}
>
Q_{0.99}\!\left(\Delta\mathrm{VSTOXX}\right)
\right]
\label{eq:method_proxy_event}
\end{equation}
Here, \(\mathrm{RV}_t\) is realised variance in window \(t\), \(d(t)\) identifies its
trading day, \(\Delta\mathrm{VSTOXX}_{d(t)}\) is the corresponding daily VSTOXX change,
and \(Q_{0.99}\) denotes the empirical 99th percentile. Thus,
\(Y_t^{\mathrm{proxy}}=1\) when either the window exhibits extreme realised variance or
its trading day belongs to the largest 1\% of VSTOXX changes.

We use this proxy only to test whether the combined anomaly score and clean-state gate identify
economically unusual conditions. Pointwise metrics assess classification, while merged episodes
evaluate coverage, false events, overlap, and lead time. VSTOXX remains external to covariance
estimation and gate construction. Because the realised-variance component favours
volatility-sensitive scores, the proxy measures economic association rather than ground-truth
anomaly accuracy. Results are reported in Section~\linkref{sec:results:anomaly_validation}.

\subsection{Detector overlap and gate validation}
\label{subsec:method_detector_validation}

For detector score \(x\), we measure separation between flagged windows (\(g=1\)) and
unflagged windows (\(g=0\)) using Cohen's \(d\),
\begin{equation}
d_x
=
\frac{\bar{x}_1-\bar{x}_0}{s_{\mathrm{pooled}}},
\qquad
s_{\mathrm{pooled}}
=
\sqrt{
\frac{(n_1-1)s_1^2+(n_0-1)s_0^2}
{n_1+n_0-2}
}
\label{eq:method_cohen_detector}
\end{equation}
Here, \(\bar{x}_g\), \(s_g^2\), and \(n_g\) are the detector-score mean, variance,
and number of observations in group \(g\). Positive values indicate that flagged windows
receive higher anomaly scores. We apply the same calculation to manifold velocity using
gate membership as the grouping variable; values near zero indicate that velocity is
largely independent of exclusion.

Detector complementarity is measured by the Jaccard coefficient between the sets of MDI- and
Isolation-Forest-flagged windows at duration \(h\). A low coefficient indicates that the
detectors identify different observations, supporting their joint use in the clean-state gate.

We also evaluate the economic ordering of the combined anomaly score. Windows are
assigned to score deciles, and we calculate mean next-window realised volatility within
each decile. Monotonicity is measured by the Spearman correlation between decile rank and
these conditional means. Tail concentration is measured by
\begin{equation}
\mathrm{Lift}_h
=
\frac{
\Pr\!\left(
\mathrm{RV}_{t+1}>Q_{0.90}(\mathrm{RV})
\mid
S^{\mathrm{comb}}_{h,t}\in\text{decile }10
\right)
}
{0.10}
\label{eq:method_anomaly_tail_lift}
\end{equation}
A lift above one indicates that the highest anomaly-score decile concentrates subsequent
volatility-tail observations. At 30 seconds, we additionally compare every gating
component and their union with the independently validated reference implementation. The
coarser grids are validated through native-construction checks and rate stability because
direct rowwise comparison across different timestamp grids is invalid.

\subsection{Alternative anomaly-handling strategies}
\label{subsec:method_anomaly_handling}

M1-M6 identify or diagnose unusual windows, whereas Strategies A-E determine how those
observations are treated in the subsequent state analysis. We apply these strategies
because hard exclusion could alter the estimated structure by removing observations
rather than modelling their anomaly severity.

\begin{table}
\tbl{Alternative treatments of identified anomalies. Strategies A-E are applied after
M1-M6 to test whether downstream state structure depends on hard exclusion. All
experiments use \(K=4\), random seed 42, and a common sample of up to 200,000 windows.
Strategy C is LE-only; Strategy E is a shared tangent-space comparator.}{
\begin{tabular}{p{2.8cm}p{4.2cm}p{4.4cm}p{1.8cm}}
\toprule
Strategy & Treatment & Robustness question & Scope\\
\midrule

A: Hard selection
& Fits the state representation using only windows with \(C_{h,t}=1\).
& Does the canonical clean-window specification produce the inferred structure?
& LE and AI\\

B: Soft weighting
& Retains all windows through anomaly-weighted resampling; flagged windows are
downweighted and calendar exclusions receive zero weight.
& Are the conclusions sensitive to complete removal rather than gradual downweighting?
& LE and AI\\

C: Score feature
& Appends the maximum daily robust \(z\)-score across MDI, Isolation Forest, and velocity
to the LE representation.
& Does anomaly severity contain useful state information?
& LE only\\

D: Separate state
& Fits \(K\) states on clean windows and assigns flagged windows to an additional anomaly
state.
& Are flagged observations better represented as a distinct state?
& LE and AI\\

E: Gaussian mixture
& Fits a regularised full-covariance Gaussian mixture to the standardised tangent
representation.
& Do the conclusions depend on the clustering model?
& Shared comparator\\

\bottomrule
\end{tabular}}
\label{tab:anomaly_handling}
\end{table}

Strategy A uses only clean windows and provides the canonical benchmark. Strategy B
retains all windows through anomaly-weighted resampling, assigning full weight to ordinary
windows, lower weights to MDI- or Isolation-Forest-flagged windows, and zero weight to
calendar exclusions. Strategy C appends the maximum daily robust \(z\)-score across MDI,
Isolation Forest, and manifold velocity to the Log-Euclidean representation. Strategy D
assigns flagged windows to a separate anomaly state, while Strategy E fits a regularised
full-covariance Gaussian mixture to test sensitivity to the clustering model. Strategy C
is restricted to Log-Euclidean geometry because an appended scalar feature has no direct
affine-invariant analogue.

For each duration and metric, we compare the economic NMI obtained under the applicable
strategies and define handling sensitivity as
\begin{equation}
\Delta_{\mathrm{NMI},h}
=
\max_{r\in\mathcal R_h}\mathrm{NMI}_{r,h}
-
\min_{r\in\mathcal R_h}\mathrm{NMI}_{r,h}
\label{eq:method_anomaly_handling_nmi}
\end{equation}
where \(r\) indexes the applicable anomaly-handling strategies. Economic NMI measures
agreement between the resulting state labels and VSTOXX tertiles. A spread below the
predeclared \(0.05\) threshold indicates that anomaly treatment does not materially
change this economic association.

\section{Continuum, Hierarchy, and State-Dynamics Methodology}\label{sec:continuum}\linkdest{sec:continuum}

This section tests the central hypothesis that masked covariance states form a connected
continuum rather than distinct regimes. We begin with discrete partitions and progressively
finer hierarchical states, examining whether they produce persistent geometric separation
and stable out-of-sample transfer. Their failure to reveal persistent boundaries motivates
formal geometric, distributional, topological, spectral, and bootstrap tests of the
continuum hypothesis. We subsequently construct a continuous stress index, separate its
slow and fast variation, estimate its relaxation dynamics, and evaluate online change
detection. All analyses use the duration-specific clean covariance clouds from Section~5
and are repeated on the native 30, 45 and 60-second grids. Affine-invariant geometry is
primary for metric-sensitive calculations, while Log-Euclidean geometry provides the
operational tangent representation and robustness comparison described in Section~4.

\subsection{Competing state hypotheses and falsification design}\label{subsec:method_hypotheses}

We distinguish between discrete and continuous representations of covariance-state
dynamics. Under the discrete-state hypothesis,
\begin{equation}
H_{\mathrm D}:\qquad
\mathcal P(\Sigma)
=
\sum_{k=1}^{K}\pi_k\mathcal P_k,
\qquad
\operatorname{sep}(\mathcal P_k,\mathcal P_\ell)>0
\quad(k\neq\ell)
\label{eq:method_hd}
\end{equation}
where \(\pi_k\) is the mixture weight of state \(k\), \(\mathcal P_k\) is its
covariance-state distribution, and positive separation implies low-density regions
between states. This hypothesis predicts multimodality, disconnected geometric
components, metastable transfer dynamics, and reproducible partitions at a meaningful
resolution \(K\).

Under the competing continuum hypothesis,
\begin{equation}
H_{\mathrm C}:\qquad
\Sigma_t=\Gamma(s_t,\xi_t),
\qquad
s_t\in\mathbb R,
\qquad
\dim_{\mathrm{eff}}(\xi_t)\ll120
\label{eq:method_hc}
\end{equation}
where \(s_t\) describes movement along a dominant state axis and \(\xi_t\) captures
lower-dimensional transverse variation. Finite regime labels are then
resolution-dependent subdivisions of a connected state space rather than naturally
separated categories.

We distinguish between the hypotheses by attempting to falsify the continuum
representation. Six predeclared certificates search for high dimensionality,
multimodality, disconnected topology, metastable spectral structure, and stable fine
partitions that would support discrete regimes. Nonlinear, residual, and year-specific
checks assess whether the conclusion persists across samples, while affine-invariant and
Log-Euclidean calculations test its metric robustness. If the continuum survives these
challenges, we represent it through a continuous stress index and use increasingly
granular hierarchical partitions to determine whether any discrete subdivision remains
stable out of sample. Slow-fast decomposition, relaxation analysis, and online
monitoring then characterise movement along this continuum.

\subsection{Six-certificate continuum evaluation framework}\label{subsec:method_certificates}

The evaluation separates discriminating from supporting evidence. Simulation calibration
identifies the dip certificate as the primary test of discreteness; dominant-axis
concentration describes low-dimensional structure, while topology, transition dynamics,
and partition stability provide supporting checks.

\paragraph{C1: Effective dimension and dominant-axis concentration}
Let \(\lambda_1\geq\cdots\geq\lambda_{120}\) denote the eigenvalues of the sample
covariance of the unscaled tangent vectors \(\phi(\Sigma_t)\). We calculate
\begin{equation}
d_{\mathrm{eff}}
=
\frac{\left(\sum_j\lambda_j\right)^2}
{\sum_j\lambda_j^2},
\qquad
f_1
=
\frac{\lambda_1}{\sum_j\lambda_j}
\label{eq:method_deff}
\end{equation}
The participation ratio \(d_{\mathrm{eff}}\) measures the effective number of active
directions, while \(f_1\) measures concentration on the dominant axis. The continuum
criterion is \(d_{\mathrm{eff}}<3\), meaning that total covariance-state variation is
concentrated in the equivalent of fewer than three equally important directions. Values
at or above three fail this certificate but do not alone establish discrete regimes. This
calculation uses the raw tangent vectors because coordinate-wise standardisation would
change their Log-Euclidean geometry. This test detects a thin support but cannot by itself
distinguish a ridge from two compact groups.

\paragraph{C2: Unimodality}
Hartigan's dip statistic is calculated on the leading principal coordinate using a
deterministic 50,000-state subsample. The criterion is
\begin{equation}
D<0.01
\label{eq:method_dip}
\end{equation}
The decision uses dip magnitude rather than its large-sample \(p\)-value, which can reject
economically negligible deviations when the number of windows is very large. A pronounced
low-density gap along the dominant coordinate would violate this criterion. Across ten
simulation replications, a genuine continuum passes with a dip statistic near 0.001,
whereas a discrete three-cluster alternative fails with a statistic near 0.126.

\paragraph{C3: \(H_0\) connected-component topology}
Zero-dimensional persistent homology tracks how connected components merge as the
distance scale increases. For 3,000 sampled states, single-linkage edges are added in
increasing order and union-find records the finite component-death scales \(w_j\). The
implemented criterion is
\begin{equation}
\frac{\max_j w_j}
{\operatorname{median}_j w_j}
>3
\label{eq:method_h0}
\end{equation}
This condition identifies one dominant connected body into which small sampling
components merge, rather than several comparably persistent islands. Thus, \(H_0\)
evaluates component separation and connectedness rather than the number of labels
generated by a clustering algorithm. This is a topological persistence calculation in
the persistent-homology sense.

\paragraph{C4: \(H_1\) loop topology}
One-dimensional persistent homology tests whether the state space contains a material
loop or ring-like structure. Let \(\ell_1\) denote the longest \(H_1\) lifetime and
\(w_0\) the largest finite \(H_0\) death scale. A loop is classified as material only when
\begin{equation}
\frac{\ell_1}{w_0}\geq0.15
\qquad\text{and}\qquad
p_{\mathrm{rank}}<0.05
\label{eq:method_h1}
\end{equation}
where \(p_{\mathrm{rank}}\) compares the observed lifetime with 20 covariance-matched
Gaussian null clouds. The continuum vote is the negation of this conjunction. This test
distinguishes a simple ridge from a ring and prevents short sampling-induced cycles from
being interpreted as economic structure.

\paragraph{C5: Transfer-operator spectral structure}
For each metric, the covariance cloud is discretised into \(M=100\) microcells. Within-day
lag-one transitions between cells define a row-stochastic matrix \(P\). If
\[
1=|\mu_1|\geq|\mu_2|\geq\cdots
\]
are its eigenvalue magnitudes, the low-index spectral-gap statistic is
\begin{equation}
G
=
\max_{1\leq j<\min(12,M)}
\frac{|\mu_j|-|\mu_{j+1}|}
{\max(|\mu_j|,10^{-9})}
\label{eq:method_spectral_gap}
\end{equation}
The continuum criterion is \(G<0.30\). Several metastable regimes would generate a small
number of near-unit eigenvalues followed by a conspicuous gap. A smoother spectrum instead
supports gradual movement through state space.

\paragraph{C6: Day-block bootstrap stability}
Partition stability is evaluated for \(k=2,\ldots,8\) using 20 bootstrap replications of
20,000 states. Each replication resamples complete trading days, refits the partition, and
compares its assignments with the reference partition using the adjusted Rand index.
If \(\bar A_k\) is the mean bootstrap ARI, the continuum criterion is
\begin{equation}
\arg\max_k\bar A_k\leq3
\qquad\text{and}\qquad
\bar A_2-\bar A_8>0.05
\label{eq:method_bootstrap}
\end{equation}
Whole-day resampling preserves intraday dependence. Stable fine partitions would support
reproducible discrete states, whereas stability concentrated at coarse resolutions is
consistent with arbitrary slicing of a connected ridge.

Certificate tallies summarise robustness rather than classify the state space. The
calibrated dip certificate is primary; dominant-axis concentration characterises the
low-dimensional structure, while effective dimension, topology, spectral structure, and
bootstrap stability provide supporting or qualifying evidence. AI remains primary, with
compatible LE evidence required for metric robustness.

\subsection{Representation robustness through the seventeen-method zoo}

The six-certificate evaluation framework assesses whether the covariance cloud supports a continuum
through effective dimension, unimodality, connected topology, loop structure, metastable spectral
gaps, and partition stability. The method zoo instead provides an independent robustness analysis
by testing whether alternative statistical, geometric, and nonlinear representations recover a
similar economic structure, while a separate Log-Euclidean transfer check evaluates sensitivity
to the sample period. It neither selects a winning method nor contributes a vote to the continuum
verdict.

The method zoo compares exactly 17 representation families: Log-Euclidean \(k\)-means;
affine-invariant Karcher \(k\)-means; diagonal Gaussian mixtures; fuzzy \(c\)-means; Ward
hierarchical clustering; spectral clustering; DBSCAN; Gaussian hidden Markov modelling;
a penalised jump model; archetypal analysis; three-component PCA followed by
\(k\)-means; scalar PC1; time-lagged independent component analysis; diffusion maps; a
PCCA+-style representation; a variational-autoencoder latent representation; and an
SPDNet-style latent representation. The comparison therefore spans hard and soft
partitions, temporal models, continuous coordinates, nonlinear embeddings, and
geometry-aware methods.

All methods use the same deterministic 8,000-state subsample from each native covariance
cloud, drawn without replacement. This controls sample differences and makes quadratic graph
and distance calculations feasible. Existing LE and AI assignments are restricted to the
selected observations, while all other representations are fitted to the same subsample.
Partition methods use \(K=4\) and \(K=8\) as comparison resolutions rather than natural
regime counts; DBSCAN selects its own groups, while PC1 and TICA remain continuous.

Hard partitions are evaluated externally against equal-frequency VSTOXX tertiles using
normalised mutual information between the inferred labels and VSTOXX categories. Continuous
coordinates are evaluated using
\begin{equation}
\left|
\operatorname{corr}(s_t,\mathrm{VSTOXX}_t)
\right|
\label{eq:method_zoo_corr}
\end{equation}
VSTOXX remains external to representation estimation and is used only to measure economic
alignment. Cross-method adjusted Rand indices quantify agreement among compatible
fixed-\(K\) partitions. DBSCAN is excluded from this agreement matrix because its noise
labels and endogenous group count are not directly comparable with fixed-\(K\)
assignments.

A separate chronological transfer exercise fits the \(K=4\) LE \(k\)-means
representation on earlier years, freezes its centroids, and evaluates its economic
alignment in later years. This transfer result applies specifically to the LE reference
partition rather than to every method in the zoo.

\label{subsec:method_continuum_checks}

We complement the method-zoo comparison with several targeted falsification checks.
First, each calendar year is treated as a separate state cloud and reassessed using
effective dimension, unimodality, and transfer-operator spectral structure. A year is
compatible with the compact continuum rule only when
\[
d_{\mathrm{eff}}<3,
\qquad
D<0.01,
\qquad
G<0.30
\]
This test determines whether the continuum interpretation persists through temporal
migration rather than being created by pooling four heterogeneous years.

Second, an adaptive diffusion map examines nonlinear geometry using
\[
W_{ij}
=
\exp\!\left[
-\frac{d_{ij}^{\,2}}{\sigma_i\sigma_j}
\right],
\]
where \(\sigma_i\) is the distance from state \(i\) to its seventh-nearest neighbour.
The calculation uses the largest connected
component of a 6,000-state subsample. A post-third-coordinate eigengap below \(0.05\)
indicates no pronounced low-dimensional split. Local bandwidths and the
connected-component restriction prevent sparse regions of the covariance cloud from
producing a degenerate diffusion spectrum.

Third, a PCCA+-style diagnostic examines whether transfer-operator memberships are soft
or nearly discrete. The diagnostic uses 100 microcells and requires both a post-\(K=4\)
eigengap below \(0.15\) and membership crispness below \(0.90\). Because the
implementation evaluates spectral separation and crispness rather than performing a
complete fuzzy-membership optimisation, it is interpreted as a PCCA+-style robustness
check.

Fourth, the no-hidden-modes test removes the leading principal direction and applies the
unimodality criterion to the dominant residual coordinate. An increase in residual
effective dimension is expected after removing PC1 and does not itself indicate hidden
regimes. The relevant falsification signal is a multimodal residual coordinate, which
would reveal discrete structure concealed by the dominant scale axis.

Finally, a granularity sweep fits partitions for \(k=2,\ldots,12\) and examines the
relative second difference of the inertia curve. A maximum relative curvature below
\(0.5\) indicates smooth improvement as additional groups are introduced rather than a
distinct elbow identifying a natural regime count.

Together, these analyses address a broader question than the evaluation framework alone. The
certificates determine whether the continuum hypothesis survives formal attempts to
reject it, while the method zoo and extended checks determine whether that conclusion
remains credible across representation families, nonlinear geometry, residual structure,
temporal subsamples, and alternative partition granularities.

\subsection{Stress-index construction, orientation, and axis stability}\label{subsec:method_dial}

For the discrete-versus-continuous regime analysis, the continuum certificates retain the
unscaled Log-Euclidean tangent isometry
\(\phi(\Sigma_t)\), ensuring that Euclidean distances equal Log-Euclidean distances. The
stress index instead uses standardised coordinates,
\begin{align}
Z_t&=\operatorname{diag}(\hat{\sigma})^{-1}
\{\phi(\Sigma_t)-\hat{\mu}\}
\label{eq:method_dial_scale}\\
s_t&=Z_tu_1
\label{eq:method_stress_dial}
\end{align}
where \(\hat{\mu}\) and \(\hat{\sigma}\) are the coordinate-wise location and scale
estimates and \(u_1\) is the leading direction of \(Z_t\). Standardisation balances the
120 coordinates when constructing the one-dimensional index but no longer preserves the
exact isometry. Its variance concentration is therefore distinct from the unscaled
dominant-axis share \(f_1\) in C1.

Because the sign of \(u_1\) is arbitrary, we orient it so that higher index values
correspond to higher VSTOXX,
\begin{equation}
u_1\leftarrow
\operatorname{sign}\!\left\{
\operatorname{corr}(Z_tu_1,\mathrm{VSTOXX}_t)
\right\}u_1
\label{eq:method_stress_orientation}
\end{equation}
VSTOXX only identifies the sign; it neither estimates the axis nor enters covariance
construction. Scaling, direction, and orientation are estimated from training observations
in each expanding-year fold. The continuum certificates (descriptive, not forecasts) use
the full masked cloud.

The operational stress index is derived from the standardised Log-Euclidean tangent
representation. Affine-invariant hierarchy calculations instead operate directly on
reconstructed covariance matrices, while forecasting and causal robustness specifications
may use a separately estimated Karcher-centred principal-geodesic coordinate. These
constructions are identified separately and are not treated as interchangeable.

We distinguish stability of the stress direction from migration in its level.
For adjacent years \(y\) and \(y+1\), directional stability is measured by
\begin{equation}
\kappa_y
=
\left|
\left(u_1^{(y)}\right)^\top u_1^{(y+1)}
\right|
\label{eq:method_axis_stability}
\end{equation}
The absolute value removes arbitrary sign reversals: values near one indicate that
successive years share a similar covariance-state direction, while values near zero
indicate substantial rotation. Year-specific means, quantiles, and densities of \(s_t\)
measure migration along that direction. This distinction allows the dominant axis to
remain stable even as the market occupies different stress levels over time.

\subsection{Hierarchical discretisation of the continuum and intraday residual structure}
\label{subsec:method_bands}\linkdest{subsec:method_bands}

Discrete labels are treated as nested approximations of the connected state cloud rather
than evidence of latent mixture components. Conceptually, thresholds
\(-\infty=c_0<c_1<\cdots<c_K=\infty\) define ordered stress bands
\begin{equation}
B_k=\{t:c_{k-1}<s_t\leq c_k\},
\qquad k=1,\ldots,K
\label{eq:method_hierarchical_bands}
\end{equation}
These bands illustrate contiguous regions along the stress continuum; the operational hierarchy
instead uses divisive, metric-specific geometric partitions. The macro tier contains \(K=4\)
groups obtained using \(k\)-means
on the standardised Log-Euclidean representation and Karcher \(k\)-means under
affine-invariant geometry. The groups are ordered by their mean \(s_t\), from calm to
stressed conditions, and each is divided into two macro-subgroups using the same global
representation.

The micro and micro-sub tiers operate on the causal intraday residual
\begin{equation}
r_t
=
\phi(\Sigma_t)
-
\frac{1}{n_{d,t}}
\sum_{\substack{u\leq t\\d(u)=d(t)}}
\phi(\Sigma_u)
\label{eq:method_intraday_residual}
\end{equation}
where \(d(t)\) is the trading day and \(n_{d,t}\) is the number of windows observed up to
time \(t\). The expanding mean includes only current and earlier windows and resets each
day, thereby removing the evolving within-day mean without using future observations.
Successive binary splits produce approximately 4, 8, 16, and 27-31 nested states; the
final count varies because a branch is divided only when its parent contains at least
3,000 observations. These states therefore provide increasingly granular geometric
approximations to the continuum, not distinct probability components.

Granularity is ranked by chronological transfer rather than in-sample variance explained.
Within each walk-forward fold, the scaler, hierarchy, centroids, and band-specific target
means are estimated from past trading days, and the first test day is embargoed. Let
\(R_0^2\) denote the persistence benchmark and \(R_k^2\) the same model augmented by
training-estimated information from tier \(k\). We define
\begin{equation}
\Delta R_k^2=R_k^2-R_0^2
\label{eq:method_band_transfer}
\end{equation}
Confidence intervals are obtained by resampling complete test days. A tier transfers only
when the lower endpoint of its 95\% interval exceeds zero. This converts ``how many
regimes?'' into the defensible question ``how finely can the state space be partitioned
before the resulting labels cease to transfer out of sample?''

We construct a continuous residual microstructure coordinate by fitting the
leading principal direction of \(r_t\) on the training observations and
projecting the held-out residuals onto it. This test distinguishes a failure
of fine discrete bands from an absence of continuous intraday information.
It is separate from the slow-fast scalar decomposition below.

\subsection{Slow-fast continuum dynamics and predictive timescales}

We next examine how movement along the continuum is distributed across persistent and
transitory timescales. For each clean native window \(t\), the stress index \(s_t\) is
decomposed using a causal, session-resetting EWMA:
\begin{equation}
s_t^{\mathrm{slow}}
=(1-\alpha_h)s_{t-1}^{\mathrm{slow}}+\alpha_hs_t,
\qquad
s_t^{\mathrm{fast}}=s_t-s_t^{\mathrm{slow}},
\qquad
\alpha_h=1-2^{-1/h}
\label{eq:method_slow_fast}
\end{equation}
Here, \(h\) is the EWMA half-life measured in native windows,
\(s_t^{\mathrm{slow}}\) captures persistent stress, and \(s_t^{\mathrm{fast}}\)
captures short-lived deviations from that level. The recursion resets each trading day,
preventing overnight information from entering the intraday state. We evaluate
\(h\in\{5,10,20,40,80,160,320\}\), while \(h=120\) defines the common descriptive
slow coordinate.

Movement of the complete covariance state is measured separately by
\begin{equation}
v_t^{(\mathrm{LE})}
=
\left\|\phi(\Sigma_t)-\phi(\Sigma_{t-1})\right\|_2
\label{eq:method_dial_velocity}
\end{equation}
where \(\Sigma_t\) is the current covariance state and \(\phi(\Sigma_t)\) is its
120-dimensional Log-Euclidean tangent vector. Thus, \(v_t^{(\mathrm{LE})}\)
measures geometric state movement, whereas \(s_t^{\mathrm{fast}}\) measures deviation
along the scalar stress axis. Both are calculated only between consecutive same-day
windows.

For every half-life, we estimate three nested models of next-window realised volatility,
\(y_{t+1}=\mathrm{RVol}_{t+1}\):
\[
\begin{aligned}
M_0:\quad &y_{t+1}\sim \mathrm{RVol}_t,\\
M_1:\quad &y_{t+1}\sim \mathrm{RVol}_t+s_t^{\mathrm{slow}},\\
M_2:\quad &y_{t+1}\sim
\mathrm{RVol}_t+s_t^{\mathrm{slow}}+s_t^{\mathrm{fast}}+v_t^{(\mathrm{LE})}.
\end{aligned}
\]
The persistent and short-run contributions are
\begin{equation}
\Delta R_{\mathrm{macro}}^2
=R^2(M_1)-R^2(M_0),
\qquad
\Delta R_{\mathrm{micro}}^2
=R^2(M_2)-R^2(M_1)
\label{eq:method_slow_fast_increments}
\end{equation}
Models are estimated through expanding-year folds with a one-day embargo and whole-day
bootstrap inference.

Finally, we compare
\begin{equation}
\Delta R_{\mathrm{total}}^2(h)
=
\Delta R_{\mathrm{macro}}^2(h)
+
\Delta R_{\mathrm{micro}}^2(h)
\label{eq:method_timescale_invariance}
\end{equation}
across half-lives. Stability of this total indicates that changing \(h\) reallocates one
continuous signal between slow and fast components rather than revealing different regime
systems. This interpretation is restricted to the tested half-life grid.

\subsection{Ornstein-Uhlenbeck and transfer-operator relaxation}\label{subsec:method_dynamics}

We estimate stress relaxation separately on the 30, 45 and 60-second grids using
consecutive within-day observations of the continuous stress index \(s_t\). A discrete
Ornstein-Uhlenbeck approximation provides the mean-reversion rate \(\theta\), long-run
stress level \(\mu\), and implied half-life \(\tau_{1/2}\); whole-day bootstrap intervals
test positive mean reversion, and frozen training estimates are evaluated against the
random-walk forecast \(s_{t+1}=s_t\). As a nonparametric check, \(s_t\) is divided into
12 ordered quantile cells, within-day movements estimate a transition matrix \(P\), and
its leading non-unit eigenvalue determines the dominant relaxation time. Agreement within
a factor of four between the OU and transfer-operator estimates is the predeclared
convergence criterion. This 12-cell analysis measures relaxation along the scalar stress
continuum, whereas the 100-microcell operator in C5 tests the full covariance cloud for
metastable regime separation. The corresponding formulas and variable definitions are
reported in Appendix~\linkref{app:relaxation_formulas}.

\subsection{Online state monitoring}

We evaluate whether movement along and around the continuum can be monitored causally at
the daily frequency. Daily means of the stress index \(s_t\) enter four complementary
detectors: BOCPD identifies abrupt state changes, two-sided CUSUM
captures persistent directional shifts, EWMA detects gradual movement,
and a 20-day kernel MMD identifies broader distributional changes. Daily mean covariance matrices also enter 252-day Riemannian control
charts: the LE chart measures tangent-space displacement, while the AI chart measures
affine-invariant distance from the trailing Karcher mean.

Signals are evaluated against the largest 5\% of absolute daily VSTOXX log changes using
a \(\pm1\)-trading-day tolerance. We report precision, recall, \(F_1\),
precision-recall area, and detection delay. Because thresholds are swept against the
evaluation events, these results provide comparative detector benchmarks rather than
untouched out-of-sample tests. A separate intraday BOCPD analysis of
\(s_t^{\mathrm{fast}}\) measures short-run change frequency and dwell time. These methods
monitor dynamics along or around the established continuum.

\section{Volatility Forecasting and Signal Evaluation}\label{sec:forecasting}

We evaluate whether the continuous covariance-state representation improves volatility
forecasts and over which horizons. First, the intraday experiments test whether it
improves on persistence across native durations and horizons ranging from the next window
to one hour. We then aggregate the state to the daily frequency and test whether it
improves next-day volatility forecasts beyond the HAR benchmark. At both frequencies,
prediction requires improvement on unseen observations rather than contemporaneous
association. Finally, a daily leading-signal analysis tests whether stress position and
motion precede volatility, with causal interpretation requiring evidence that survives
reverse-direction, chronological out-of-sample, and null tests.

\subsection{Forecast Targets and Eligibility}
\label{subsec:fc_objects}

Let \(t\) index a covariance window of native duration
\(\delta\in\{30,45,60\}\) seconds, and let \(\mathcal D(t)\) denote its trading
session. At the close of window \(t\), each forecast uses only contemporaneously
available limit-order-book features, regularised covariance states, and
realised-volatility observations from window \(t\) and earlier windows. The
strictly forward target is
\begingroup
\setlength{\abovedisplayskip}{5pt}
\setlength{\belowdisplayskip}{5pt}
\begin{equation}
 RV_{t,H}^{+}
 =\left[
   \sum_{\substack{u:\ t<\tau_u\leq \tau_t+H\\
                    \mathcal D(u)=\mathcal D(t)}}rv_u^2
  \right]^{1/2}
 \label{eq:fc_forward_rv}
\end{equation}
\endgroup
Here, \(\tau_u\) is the timestamp of future window \(u\), \(H\) is the forecast
horizon, and \(rv_u\) is its realised volatility. Variance increments are
accumulated before returning the target to volatility units.
A target is retained only when the
complete interval \((t,t+H]\) is observable within the anchor session and
evaluation fold; it is never shortened, carried overnight, or imputed. The
next-window target uses the next complete native covariance window.

The horizon grid includes the next native window; 30, 45 and 60 seconds; and
2, 5, 10, 15, 30 and 60 minutes. Horizons shorter than the native duration are
unavailable, and cells with less than \(20\%\) target coverage are excluded,
leaving 54 admissible combinations across three durations, two geometries and
ten candidate horizons. Held-out gains establish prediction, while lead and
causal claims additionally require temporal precedence and dedicated
out-of-sample evidence.

\subsection{Hierarchical state transfer and predictive granularity}
\label{subsec:fc_hierarchy}

We first subject the continuum representation to a hierarchical resolution test
that probes whether its dominant stress axis conceals forecast-relevant local
structure. Rather than presuming a fixed regime count, we recursively partition
the covariance cloud into macro, macro-sub, micro, and micro-substates and
evaluate each resolution through expanding-year transfer under LE and AI
geometry. We then evaluate the continuous representation through stress
position, short-run displacement, and covariance-state velocity. Improved
forecasting from finer partitions would indicate discrete structure missed by
the dominant axis, whereas declining hierarchical transfer alongside persistent
continuous predictability would show that economically useful information
survives in the covariance geometry but is lost through increasingly narrow
discretisation.

Using the hierarchy defined in Section~\linkref{subsec:method_bands}, each
expanding-year fold adds training-estimated state-level next-window volatility
to the persistence benchmark. Incremental \(R^2\) is evaluated across all
durations and geometries using 95\% intervals from 400 day-block bootstrap
replications; transfer requires the lower bound to exceed zero. Gains confined
to coarse tiers support broad stress segmentation, whereas fine-tier gains
indicate additional local structure.

\subsection{Continuous intraday predictors and fifteen-model bake-off}
\label{subsec:fc_predictors}
\label{subsec:fc_models}

Every intraday learner receives the same four-variable predictor set,
\begingroup
\setlength{\abovedisplayskip}{5pt}
\setlength{\belowdisplayskip}{5pt}
\begin{equation}
 F_t^{(m,\delta)}
 =\left[rv_t,\ s^{(m)}_{\mathrm{slow},t},\
 s^{(m)}_{\mathrm{fast},t},\ v_t^{(m)}\right],
 \qquad m\in\{\mathrm{LE},\mathrm{AI}\}
 \label{eq:fc_predictor_vector}
\end{equation}
\endgroup
Here, \(rv_t\) controls for volatility persistence and the remaining variables
represent continuous covariance-state level, short-run displacement, and
metric-native movement, as defined in Section~\linkref{sec:continuum}. Specifications
are estimated separately for each duration and geometry.

Because velocity is undefined at session openings, the forecasting array replaces these
values with zero. This convention affects approximately \(0.18\%\) of each
duration sample and is treated as an implementation limitation rather than
observed zero movement.

At each intraday horizon, this specification is evaluated against a fold-specific
realised-volatility persistence benchmark,
\begingroup
\setlength{\abovedisplayskip}{5pt}
\setlength{\belowdisplayskip}{5pt}
\begin{equation}
 RV_{t,H}^{+}=\alpha_H+\rho_Hrv_t+\varepsilon_{t,H}
 \label{eq:fc_persistence}
\end{equation}
\endgroup
estimated using the same observations as every competing model.

Functional-form robustness is examined using 15 predictive architectures: OLS,
ridge, lasso, elastic net, Huber, quadratic and cubic regression,
\(k\)-nearest neighbours, radial-basis SVR, kernel ridge, gradient boosting,
random forest, extremely randomised trees, a multilayer perceptron, and
Gaussian-process regression. This panel tests whether predictive performance
depends on linearity, regularisation, nonlinear interactions, local similarity,
kernels, trees, or neural approximation. Hyperparameters are selected through
three chronological inner splits, with the tuning sample capped at 30,000
training observations. All model-specific transformations
are estimated using training data only. The experiment evaluates continuous
predictors, while hierarchical transfer is examined separately in
Section~\linkref{subsec:method_bands}.

\subsection{Next-day HAR and state augmentation}
\label{subsec:fc_nextday}

The daily experiment tests whether intraday covariance-state information retains
predictive value after aggregation. For trading day \(d\), daily realised
volatility and the next-day target are
\begingroup
\setlength{\abovedisplayskip}{5pt}
\setlength{\belowdisplayskip}{5pt}
\begin{equation}
 RV_d=\left(\sum_{t:\mathcal D(t)=d}rv_t^2\right)^{1/2},
 \qquad Y_d=RV_{d+1}
 \label{eq:fc_daily_target}
\end{equation}
\endgroup
The benchmark HAR block contains daily, weekly, and monthly volatility
components,
\begingroup
\setlength{\abovedisplayskip}{5pt}
\setlength{\belowdisplayskip}{5pt}
\[
 H_d=\left[RV_d,\ \overline{RV}_{d,5},\
 \overline{RV}_{d,22}\right],
\]
\endgroup
where the bars denote trailing means. A separate LOB block
contains daily spread, log-Kyle price impact, log-Amihud illiquidity, and
VPIN-style order-flow toxicity. The continuum block summarizes daily stress
through its mean, maximum, closing value, standard deviation, average absolute
movement, and high-stress-band share, the only hierarchical input. Six OLS specifications
test whether this block improves on HAR and HAR-LOB using common dates.
Fifteen learners assess model capacity, while five training-fold permutations
test dependence on temporal alignment; nested OLS comparisons remain primary.

\subsection{Out-of-Sample Forecast Evaluation and Inference}
\label{subsec:ev_walkforward}
\label{subsec:ev_losses}

Forecasts are evaluated through expanding-year folds: each model is trained on
all completed earlier years and tested on the subsequent year, with the first
test day embargoed. Competing models use identical rows,
with the covariance state-coordinate representation (standardisation, PCA axis, orientation)
and model parameters all estimated from training data in each fold. Out-of-sample predictions are concatenated before ranking
to separate model selection from evaluation. Accuracy is measured using RMSE, MAE, signed bias,
and volatility-specific QLIKE loss. A positive floor prevents undefined QLIKE values; it is
estimated from the training sample only, so the evaluation convention never depends on held-out
data.

Absolute performance is measured by out-of-sample \(R^2\), while incremental \(R^2\) compares
each model with realised-volatility persistence intraday and HAR daily. Training performance and
train-test gaps serve only as capacity and overfitting diagnostics. Mincer-Zarnowitz intercepts,
slopes, regression \(R^2\), and joint Wald tests assess calibration; forecasts that reject the
calibration benchmark are classified as predictive but miscalibrated. Performance is also
evaluated across years, VSTOXX tertiles, stress levels, and future-volatility tails. The remaining
implementation qualification is zero replacement for undefined opening velocity. Reported findings
therefore apply to the executed specification.

\label{subsec:ev_inference}
\label{subsec:ev_gates}

Since intraday targets overlap, uncertainty is estimated by resampling complete trading days and
applying identical bootstrap draws across models, preserving target dependence and within-session
dynamics. Next-day comparisons use corrected Diebold-Mariano tests and the 90\% Model Confidence
Set. Performance is evaluated by year, VSTOXX tertile, stress level, and the upper decile of future
realised volatility. Intraday skill requires at least 20\% target coverage and a positive
day-bootstrap interval for incremental \(R^2\), while next-day improvement requires lower loss than
HAR, a favourable Diebold-Mariano result, and Model Confidence Set survival. Training-fold
permutations assess the next-day regime block, while broader intraday claims rely on chronological
out-of-sample tests and uncertainty intervals. Robustness requires agreement across LE and AI
geometry and the 30, 45 and 60-second grids. Because the grids are dependent views of the same
message stream, cross-duration agreement represents resolution robustness rather than independent
replication; similarly, LE-AI agreement reflects empirical metric robustness rather than metric
equivalence. Automated checks enforce timestamp, target, fold, and feature alignment.

\subsection{Leading-signal and causal falsification}
\label{subsec:ev_state_validation}
\label{subsec:ev_causal_arbiter}
\label{subsec:ev_reproducibility}

Leading-signal analysis separates stress position from within-session velocity and acceleration,
with derivatives reset at session boundaries. Event studies examine these variables, realised
volatility, and VSTOXX within \([-2,+2]\) days of the largest 5\% of absolute VSTOXX changes,
while lead-lag correlations and bidirectional Granger tests assess temporal ordering. Chronological
out-of-sample tests provide the decisive criterion: a leading effect requires a strictly positive
incremental-performance interval across durations and geometries.  In each expanding-year fold,
the covariance dial, its orientation, and the forecasting model are estimated using prior years
only.  Targets crossing a fold boundary are purged and the first test day is embargoed.
Confidence intervals use 2,000 five-day moving-block bootstrap replications within test years,
with ten-day blocks as a sensitivity check.  Event-time or Granger evidence without
out-of-sample support remains descriptive.

Critical slowing down is assessed through pre-event increases in autocorrelation and variance
relative to random non-event dates, with evidence required to exceed this benchmark across
specifications. These tests are evaluated separately rather than by majority vote. Each run records
its design, sample, parameters, skipped cells, and software configuration. Arrays are aligned by
timestamp and session, with hard failure on mismatches. Leakage checks cover target timing, fold
boundaries, training-only transformations, embargoes, comparison rows, and resampling units.
Reported findings include only executed analyses.

\section{Empirical Results}\label{sec:results}

\subsection{Covariance-State Foundation and Metric Robustness}\label{sec:results:sample}

\begin{figure}[t]
\centering
\includegraphics[width=0.98\linewidth]{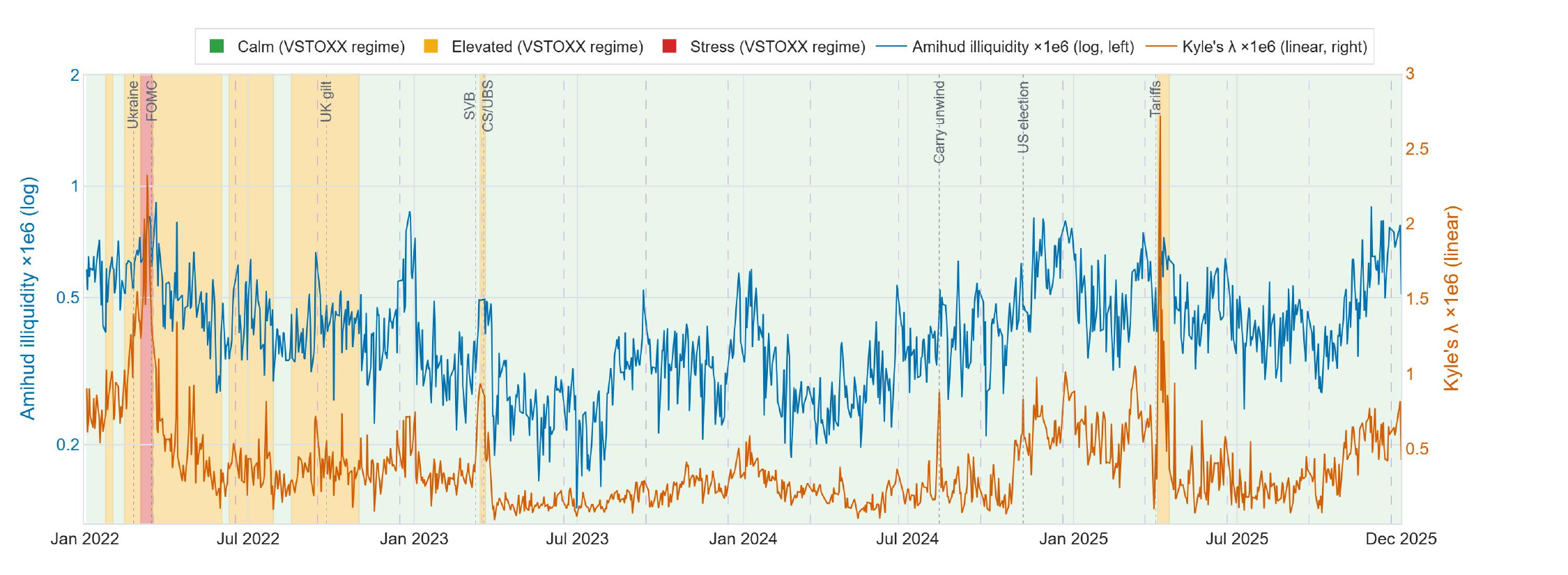}
\caption{Daily Amihud illiquidity and Kyle price impact across VSTOXX regimes,
2022-2025.  Amihud illiquidity is shown on the logarithmic left axis and
Kyle's lambda on the linear right axis; shaded bands identify calm, elevated,
and stress regimes.  Observations within one calendar day of FESX expiry/roll
dates are excluded.}
\label{fig:amihud_kyle_daily}
\end{figure}

\begin{figure}[t]
\centering
\includegraphics[width=0.72\linewidth]{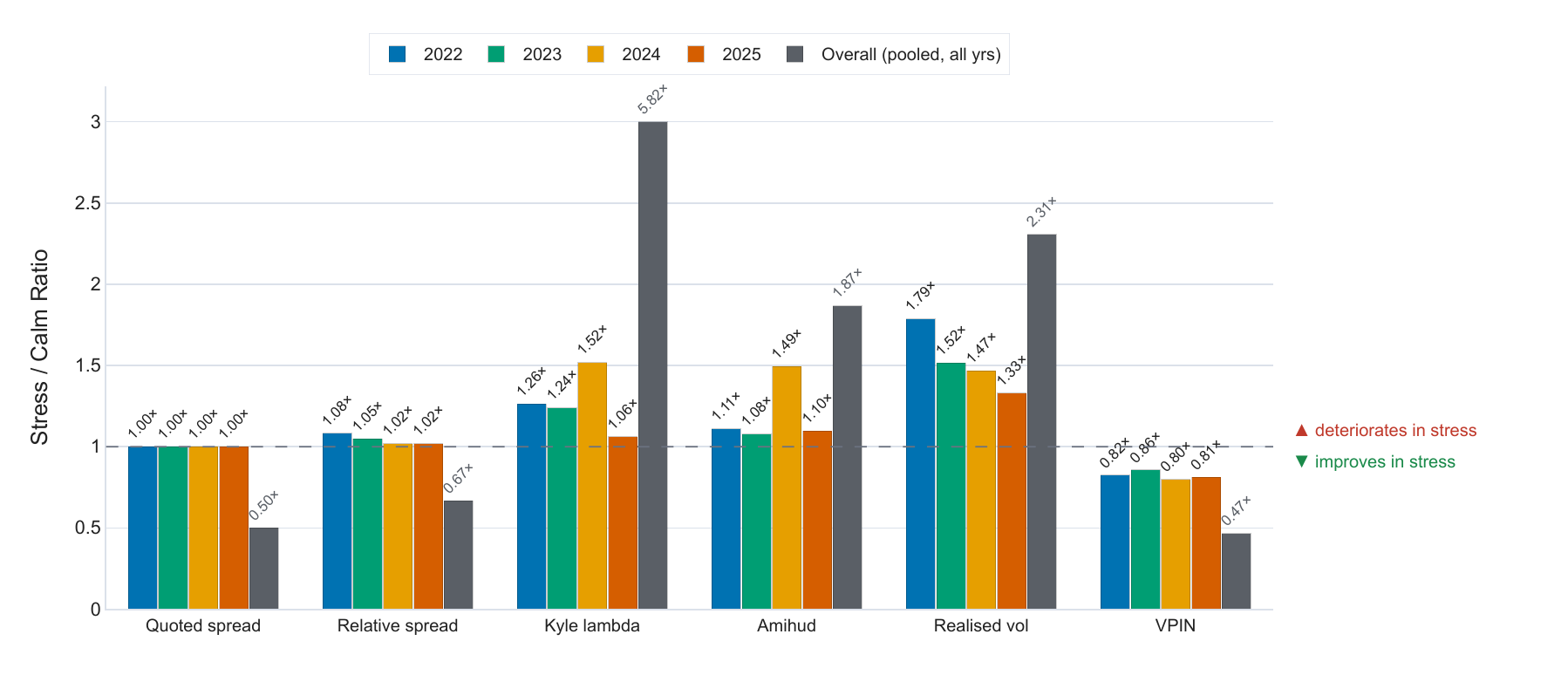}
\caption{Stress-to-calm ratios for spreads, price impact, illiquidity, realised
volatility, and VPIN.  Bars show within-year VSTOXX terciles and pooled
estimates; the dashed line marks equal medians.}
\label{fig:spread_lies_yearwise}
\end{figure}

The canonical analysis spans 987 trading days and contains 837,890, 559,261,
and 419,493 clean covariance windows on the native 30, 45 and 60-second
grids.  Relative to their unmasked counterparts, the gate retains 80.71\%,
80.76\%, and 80.81\%.  The near equality is useful: duration changes the number
of observations, but not the economic severity of the filter.
The covariance state combines 15 event-level LOB variables covering prices,
quoted and relative spreads, microprice, displayed depth, trade volume,
multi-level imbalance, order-flow imbalance, and price pressure
(Appendix Table~\ref{tab:app_covariance_feature_inventory}).  These variables
represent complementary channels of liquidity, while their covariance captures
how trading costs, available depth, market activity, and directional pressure
move together.  Amihud illiquidity and Kyle's lambda serve as external economic
benchmarks for interpreting the feature set, with higher values indicating
greater price movement relative to trading activity and stronger price impact
from order flow, though they remain outside the covariance matrix.  As shown in
Figure~\ref{fig:amihud_kyle_daily}, both rise around the principal stress
episodes, illustrating the deterioration in price impact and market resilience
that the selected LOB variables are intended to represent.
Figure~\ref{fig:spread_lies_yearwise} further motivates the multivariate design
by showing that quoted spreads can remain stable or improve even as price
impact, illiquidity, and realised volatility worsen.  The figures therefore
provide economic context for the feature set and explain why liquidity cannot be
represented by quoted spread alone, while direct validation of the covariance
state is assessed separately through its economic ordering and predictive
performance.

The quoted-spread reversal shows why displayed spread alone provides an
incomplete view of liquidity.  In the canonical 30-second sample, mean spread
narrows from \(0.000339\) in the calmest regime to \(0.000271\) in the most
stressed, even as depth, price impact, and realised volatility deteriorate;
``spread lies'' therefore describes its partial economic signal.  Scale also
affects the geometric representation: the leading component captures most
variation in the raw metric-preserving LE coordinates but only
26.44\%-29.31\% after coordinate-wise standardisation.  Metric-foundation
acceptance requires LE-AI agreement of at least 0.90 and yearly drift
correlation of 0.80.  The 30-second sample meets both criteria, the 45-second
sample meets only the drift criterion, and the 60-second sample meets neither,
supporting AI as the primary geometry and LE as a duration- and scale-sensitive
robustness benchmark.

\clearpage
\subsection{Anomaly-Gate Validation and Clean-Sample Robustness}
\label{sec:results:anomaly_validation}\linkdest{sec:results:anomaly_validation}

\begin{figure}[!b]
\centering
\includegraphics[width=0.98\linewidth]{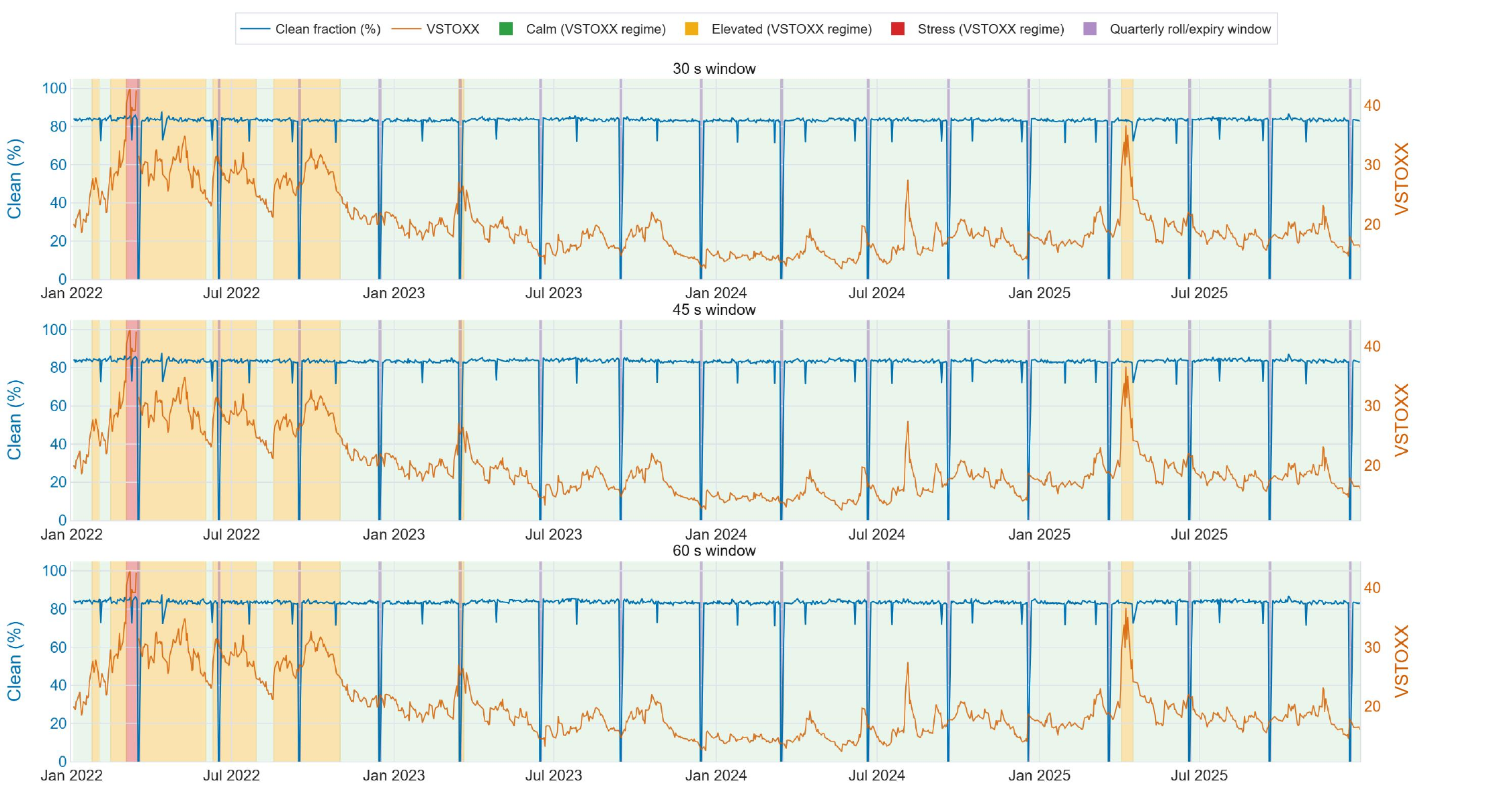}
\caption{Daily clean-window retention across the 30, 45 and 60-second grids,
with VSTOXX and roll/expiry periods.  Retention is stable across durations and
volatility regimes; sharp declines reflect predetermined exclusions.}
\label{fig:clean_fraction_native}
\end{figure}

Before comparing market-state geometry across durations, we filter atypical trajectories
and non-continuous trading periods that could distort covariance estimates or create
artificial state separation, while retaining economically meaningful market stress. The
independently constructed 30, 45 and 60-second masks each retain approximately 80.7--80.8\% of
their native windows. Figure~\ref{fig:clean_fraction_native} confirms this stability;
sharp declines are concentrated in predetermined contract-roll and expiry exclusions.
Exact reproduction of all 30-second reference
flags validates the implementation against the reference pipeline. The independently
constructed 45- and 60-second masks are validated through native-grid construction
checks, stable retention rates across years and VSTOXX regimes, and the common
progression from 1,023 processed days to 1,018 evaluated and 987 final days.
Together, these checks support the reliability and cross-resolution consistency of
the gate. Table~\ref{tab:sample_gate} provides
detector-level evidence: matrix-discord flags produce strong separation, with
Cohen's \(d\) close to two, while Isolation Forest separates flagged
observations even more sharply.  Their Jaccard overlap remains below 0.033,
indicating that temporal-trajectory anomalies and unusual multivariate
configurations capture distinct but complementary forms of market behaviour.
Manifold velocity shows little separation, supporting its use as a transition
diagnostic rather than an exclusion criterion.

The combined anomaly score provides economic evidence that the gate selects
meaningful market states rather than arbitrary observations.  Its deciles
correlate 0.952-0.988 with next-window realised volatility, and the highest
decile contains 2.56-2.71 times the unconditional incidence of
volatility-tail events.  The weaker association with VSTOXX suggests that the
score captures high-frequency order-book conditions beyond daily market-wide
volatility.  At the same time, proxy AUROC of 0.670-0.688, precision near
3.4\%, recall near 32\%, and zero median lead time show that the gate is best
interpreted as a contemporaneous state-selection mechanism rather than an
anomaly classifier or early-warning signal.  Table~\ref{tab:mask_reversal}
provides the consequential downstream test: masking restores full LE-AI
certification at 30 seconds, improves drift agreement at 45 seconds, and
weakens both geometric diagnostics at 60 seconds.  Because the masking rules
were specified in advance, we apply the same gate across durations rather than
adjusting it after observing the results, avoiding outcome-driven selection.

\begin{table}[t]
\centering
\caption{Native-grid sample, detector, and economic validation.  The economic lift compares
the realised-volatility tail probability in the highest combined-score decile
with its unconditional 10\% rate.}
\label{tab:sample_gate}
\small
\begin{tabular}{lrrr}
\toprule
 & 30 s & 45 s & 60 s\\
\midrule
Unmasked windows & 1,038,115 & 692,481 & 519,121\\
Canonical clean windows & 837,890 & 559,261 & 419,493\\
Retained fraction (\%) & 80.71 & 80.76 & 80.81\\
Matrix-profile own-flag AUROC & 0.9834 & 0.9848 & 0.9860\\
Matrix-profile Cohen's $d$ & 1.930 & 1.964 & 1.996\\
Isolation Forest Cohen's $d$ & 3.674 & 3.754 & 3.819\\
MDI-Isolation Forest Jaccard & 0.0249 & 0.0279 & 0.0326\\
Velocity-gate Cohen's $d$ & $-0.09$ & 0.02 & 0.16\\
Decile-contemporaneous-volatility Spearman correlation & 0.952 & 0.915 & 0.964\\
Decile-future-volatility Spearman correlation & 0.988 & 0.952 & 0.988\\
Decile-VSTOXX Spearman correlation & 0.733 & 0.733 & 0.867\\
Top-decile volatility-tail probability (\%) & 26.60 & 25.57 & 27.14\\
Top-decile tail lift & 2.660 & 2.557 & 2.714\\
Combined-score proxy AUROC & 0.679 & 0.670 & 0.688\\
Proxy gate precision & 0.0346 & 0.0343 & 0.0347\\
Proxy gate recall & 0.3263 & 0.3236 & 0.3285\\
\bottomrule
\end{tabular}
\begin{minipage}{0.94\linewidth}\footnotesize
Proxy metrics use an intentionally incomplete volatility/VSTOXX label and are
reported only as diagnostics.
\end{minipage}
\end{table}

\begin{table}[t]
\centering
\caption{Masked-unmasked LE/AI foundation reversal.  The operating minimum
has a 0.9 certification threshold.}
\label{tab:mask_reversal}
\small
\begin{tabular}{llrrl}
\toprule
Duration & Sample & Operating minimum & Drift correlation & Status\\
\midrule
30 s & Unmasked & 0.8778 & 0.7932 & Not certified\\
     & Masked   & 0.9532 & 0.8600 & Certified\\
45 s & Unmasked & 0.5499 & 0.4622 & Not certified\\
     & Masked   & 0.8178 & 0.8437 & Drift-only reversal\\
60 s & Unmasked & 0.9518 & 0.1493 & Operating gate passes\\
     & Masked   & 0.7518 & $-0.1666$ & Masking hurts foundation\\
\bottomrule
\end{tabular}
\begin{minipage}{0.94\linewidth}\footnotesize
The canonical sample remains the native mask at every duration; ``status''
describes only this foundation diagnostic.
\end{minipage}
\end{table}

\subsection{Continuum Certificates, Falsification, and Metric Robustness}
\label{sec:results:continuum}

The central question is whether covariance states exhibit reproducible
boundaries that justify discrete regimes or instead form a connected stress
continuum. To avoid relying on any single statistic or geometry, we assess
this question using six complementary dimensional, distributional,
topological, spectral, and resampling tests across all three covariance
durations and both SPD metrics. Table~\ref{tab:continuum_scoreboard} provides consistent
evidence for a continuum. The consistently low effective dimension, strong
leading component, negligible multimodality, and weak topological persistence
provide no evidence of separated regimes. The leading axis also correlates
with VSTOXX, although VSTOXX is excluded from its estimation, giving the
stress coordinate an independent economic interpretation. Its modest
strengthening at longer durations suggests that temporal averaging reduces
local noise without altering the underlying ordering of market states.
Together, these results support a continuous, economically meaningful
representation of covariance states and provide the foundation for the
analysis that follows.

\begin{figure}[t]
\centering
\includegraphics[width=0.98\linewidth]{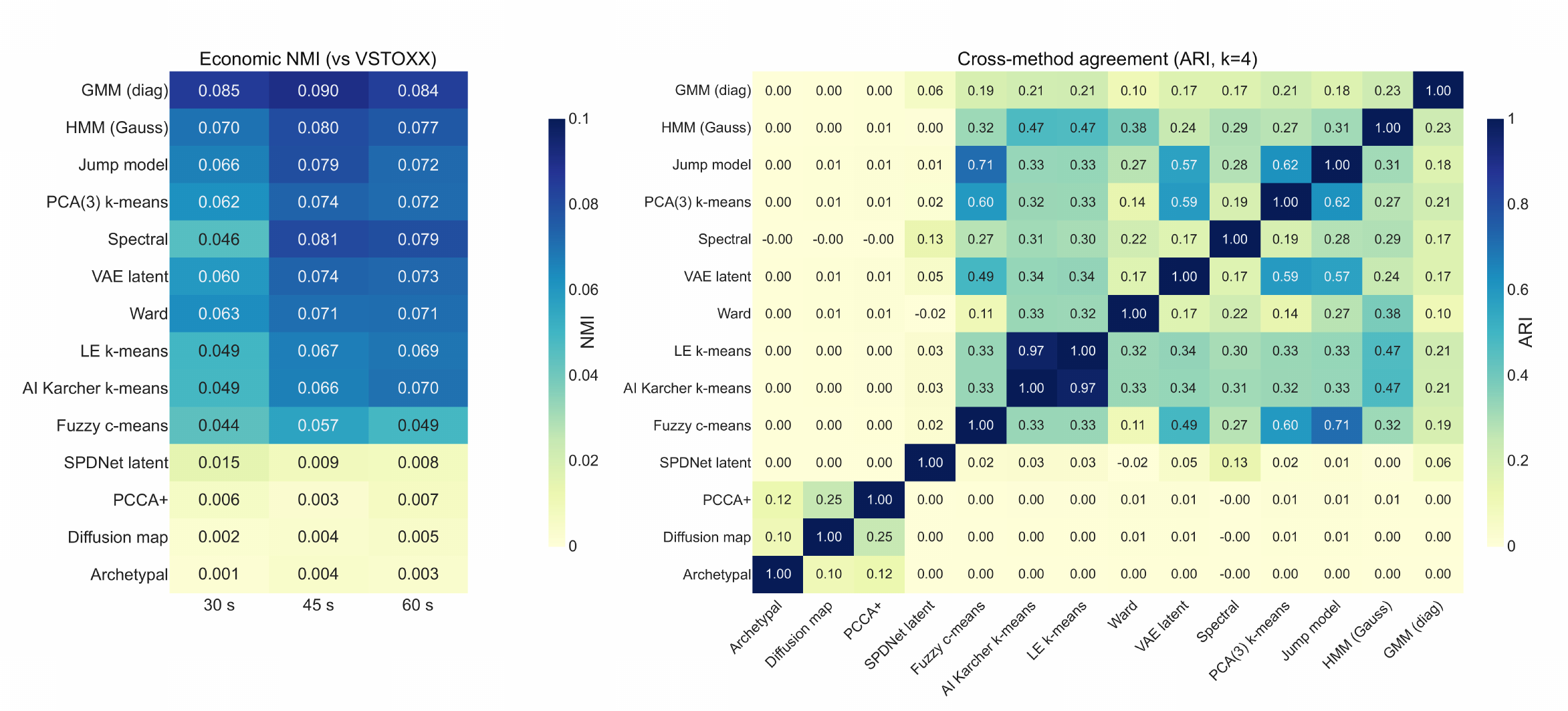}
\caption{Method-zoo robustness across clustering and embedding approaches.
The left panel reports NMI between masked four-state partitions and VSTOXX
tertiles at each covariance duration; the right panel reports pairwise ARI
across methods for the 30-second sample.}
\label{fig:method_zoo_robustness}
\end{figure}

Table~\ref{tab:continuum_scoreboard} and
Figure~\ref{fig:method_zoo_robustness} show that the evidence for a continuum is
robust to covariance duration, SPD metric, and clustering method. The
calibrated unimodality certificate provides the primary distinction: simulated
dip statistics are approximately 0.001 for a genuine continuum and 0.126 for a
discrete three-cluster alternative. Observed values range from 0.00070 to 0.00096,
while PC1 explains 83.66\%-85.27\% of covariance-state variation. The remaining
diagnostics support this connected, low-dimensional interpretation, with the
45-second AI bootstrap result providing a qualification.
Figure~\ref{fig:method_zoo_robustness} tests whether alternative clustering
methods recover stable and economically meaningful regimes. They do not:
VSTOXX NMI is at most 0.072, indicating only weak alignment between the hard
clusters and the external stress categories, while DBSCAN identifies no
natural separation and collapses the sample into one group. The LE and AI
$K=4$ partitions nevertheless agree strongly (ARI $=0.969$), showing that the
two geometries divide the data similarly when four clusters are imposed.
Thus, the partition is reproducible across metrics but has weak external
economic separation, and its exact boundaries remain method-dependent. By
contrast, the continuous PC1 coordinate retains an absolute VSTOXX
correlation of 0.488 on the common sample, supporting a stable continuous
stress ordering rather than a uniquely identifiable set of discrete regimes.

\begin{table}[t]
\centering
\caption{Continuum evidence, certificate scoreboard, and metric robustness.}
\label{tab:continuum_scoreboard}
\scriptsize
\setlength{\tabcolsep}{3.5pt}
\begin{tabular}{llccc}
\toprule
Certificate or measure & Acceptance rule & 30 s & 45 s & 60 s\\
\midrule
Clean windows & - & 837,890 & 559,261 & 419,493\\
Raw PC1 variance (\%) & - & 83.66 & 84.41 & 85.27\\
$|\mathrm{corr}(\mathrm{PC1},\mathrm{VSTOXX})|$ &
External validation & 0.500 & 0.514 & 0.522\\
\midrule
C1: Effective dimension & $d_{\mathrm{eff}}<3$ &
1.4109 (pass) & 1.3887 (pass) & 1.3634 (pass)\\
C2: PC1 unimodality & dip $<0.01$ &
0.00080 (pass) & 0.00070 (pass) & 0.00096 (pass)\\
C3: $H_0$ connectedness & death ratio $>3$ &
43.579 (pass) & 26.084 (pass) & 25.741 (pass)\\
C4: No persistent $H_1$ loop & ratio $<0.15$ or rank $p\geq0.05$ &
0.0209 (pass) & 0.0568 (pass) & 0.0437 (pass)\\
C5: Spectral gap, AI/LE & each $<0.30$ &
0.176/0.227 (pass) & 0.139/0.200 (pass) & 0.188/0.187 (pass)\\
C6: Bootstrap peak $K$, AI/LE & each $K\leq3$; ARI drop $>0.05$ &
2/2 (pass) & 4/2 (AI fail) & 3/2 (pass)\\
\midrule
Diagnostic tally, AI/LE & descriptive summary &
6/6; 6/6 & 5/6; 6/6 & 6/6; 6/6\\
Overall assessment & calibrated and supporting evidence &
Supported & Supported, qualified & Supported\\
\bottomrule
\end{tabular}
\begin{minipage}{0.94\linewidth}\footnotesize
Paired entries report AI/LE values. The tally summarises all diagnostics; the
calibrated dip certificate is the primary discriminator, and the 45-second AI
bootstrap result qualifies partition stability.
\end{minipage}
\end{table}

Year-specific tests confirm that the pooled continuum does not conceal a
structural break. All 45- and 60-second annual samples pass, as do the
30-second samples for 2022-2024. The only qualifications are marginal
30-second spectral-gap failures for 2025 (0.3103) and the 2024-2025 forward
split (0.3034), both just above the 0.30 threshold. The 2025 cloud nevertheless
remains low-dimensional and unimodal (effective dimension 1.6645, PC1 variance
76.05\%, dip 0.000655, and absolute VSTOXX correlation 0.748). Removing PC1
also reveals no hidden mode: residual dimension increases, but the PC2 dip
remains only 0.00064-0.00078. These results support a stable continuum rather
than discrete regimes; complete year-specific diagnostics are reported in the
supplementary materials.

The final robustness check evaluates whether the stress axis changes over
time. Adjacent-year PC1 cosines remain between 0.974 and 0.998, showing that
its direction is highly stable, while its oriented association with VSTOXX
strengthens from approximately 0.42 in 2022 to 0.75-0.77 in 2025. The market
therefore moves across different levels of a stable stress axis rather than
shifting to a new underlying structure. Because these yearwise folds are
dependent subperiods of one instrument, they should be interpreted as
chronological stress tests rather than independent replications. Together,
the geometric certificates, method-zoo comparison, year-specific tests, and
hidden-mode checks support a stable, low-dimensional, and economically
meaningful continuum across durations and SPD metrics. This result underpins
the remaining analysis: discrete regimes are treated as descriptive level
sets along the continuous stress axis rather than as naturally separated
market states.

\subsection{Anomaly Sensitivity and Hawkes Robustness}
\label{sec:results:alternative_anomaly_hawkes}

A credible continuum finding must remain stable across alternative treatments
of anomalous windows. Tables~\ref{tab:alternative_anomaly_strategies}
and~\ref{tab:anomaly_hawkes_robustness} therefore compare five approaches:
binary deletion, soft weighting, inclusion of the anomaly score as an
additional LE coordinate, assignment to a separate anomaly regime, and robust
Gaussian-mixture clustering. Strategy C has no AI analogue because it appends
a Euclidean coordinate. Robust GMM produces the highest economic NMI in five
of the six duration-metric comparisons; binary deletion leads only for
60-second LE, by 0.0009, while the separate anomaly regime consistently has
the weakest alignment. More importantly, the spread across treatments remains
below the pre-specified 0.05 materiality threshold at every duration: 0.0298,
0.0300, and 0.0230 for LE and 0.0380, 0.0272, and 0.0224 for AI at 30, 45,
and 60 seconds. The continuum conclusion is therefore robust to whether
anomalous observations are deleted, down-weighted, represented through their
score, or isolated as a separate regime.

\begin{table}[!b]
\centering
\caption{Economic NMI under alternative anomaly treatments.}
\label{tab:alternative_anomaly_strategies}
\scriptsize
\setlength{\tabcolsep}{5pt}
\begin{tabular}{lccc}
\toprule
Treatment & 30 s & 45 s & 60 s\\
\midrule
A: hard deletion & 0.0611 / 0.0468 & 0.0670 / 0.0654 & 0.0767 / 0.0732\\
B: soft weighting & 0.0597 / 0.0457 & 0.0705 / 0.0624 & 0.0724 / 0.0696\\
C: score feature & 0.0576 / - & 0.0729 / - & 0.0742 / -\\
D: anomaly regime & 0.0424 / 0.0342 & 0.0449 / 0.0476 & 0.0537 / 0.0535\\
E: robust GMM & 0.0722 / 0.0722 & 0.0748 / 0.0748 & 0.0758 / 0.0758\\
\bottomrule
\end{tabular}
\begin{minipage}{0.94\linewidth}\footnotesize
Entries report LE / AI economic NMI.  Strategy C has no affine-invariant
analogue because it appends the anomaly score as a Euclidean coordinate.
\end{minipage}
\par\vspace{2\baselineskip}
\caption{Anomaly-treatment materiality and Hawkes robustness.}
\label{tab:anomaly_hawkes_robustness}
\scriptsize
\setlength{\tabcolsep}{2.5pt}
\begin{tabular}{lrrrrrrl}
\toprule
Duration & \shortstack{LE\\spread} & \shortstack{AI\\spread} &
\shortstack{Mean\\$n$} & \shortstack{$\mathrm{corr}$\\$(n,\mathrm{VSTOXX})$} &
\shortstack{LE $n$\\spread} & \shortstack{AI $n$\\spread} & Verdict\\
\midrule
30 s & 0.0298 & 0.0380 & 0.4417 & 0.1631 & 0.0318 & 0.0138 & Pass / negative\\
45 s & 0.0300 & 0.0272 & 0.4419 & 0.1627 & 0.0251 & 0.0161 & Pass / negative\\
60 s & 0.0230 & 0.0224 & 0.4418 & 0.1634 & 0.0232 & 0.0168 & Pass / negative\\
\bottomrule
\end{tabular}
\begin{minipage}{0.94\linewidth}\footnotesize
The anomaly-treatment materiality gate is an economic-NMI spread below 0.05.
The reflexivity-flat gates are a regime-$n$ spread below 0.10 and
$|\mathrm{corr}(n,\mathrm{VSTOXX})|<0.20$.  ``Pass / negative'' denotes a
materiality pass and a negative result for separable Hawkes reflexivity.
\end{minipage}
\end{table}

The Hawkes analysis evaluates whether the covariance-state continuum also
distinguishes self-exciting from externally driven stress.
Table~\ref{tab:anomaly_hawkes_robustness} shows that mean branching ratios are
subcritical and nearly constant across durations (0.4417-0.4419), with VSTOXX
correlations of approximately 0.163. Branching-ratio spreads across the four
covariance-state partitions also remain small, ranging from 0.0232 to 0.0318
under LE and from 0.0138 to 0.0168 under AI. These values satisfy the
pre-specified reflexivity-flat criteria of a regime spread below 0.10 and an
absolute VSTOXX correlation below 0.20, producing the ``Pass / negative''
verdict. Only 0.6\% of observations have branching ratios above 0.70, while
5.9\% fall below 0.30, and the branching-ratio relationship with the
highest-VSTOXX regime changes direction between LE and AI. The available
Hawkes measure therefore provides no stable separation between endogenous and
exogenous stress. This negative result reflects the absence of signed
transaction information in the snapshot trade feed rather than evidence that
stress is exogenous. Distinguishing these mechanisms remains an important
direction for future work using signed transaction-level data and richer
order-flow models.

\subsection{Discrete bands expose a hierarchy-overfit pattern}
\label{sec:results:hierarchy}\linkdest{sec:results:hierarchy}

\begin{figure}[!b]
\centering
\includegraphics[width=0.98\linewidth]{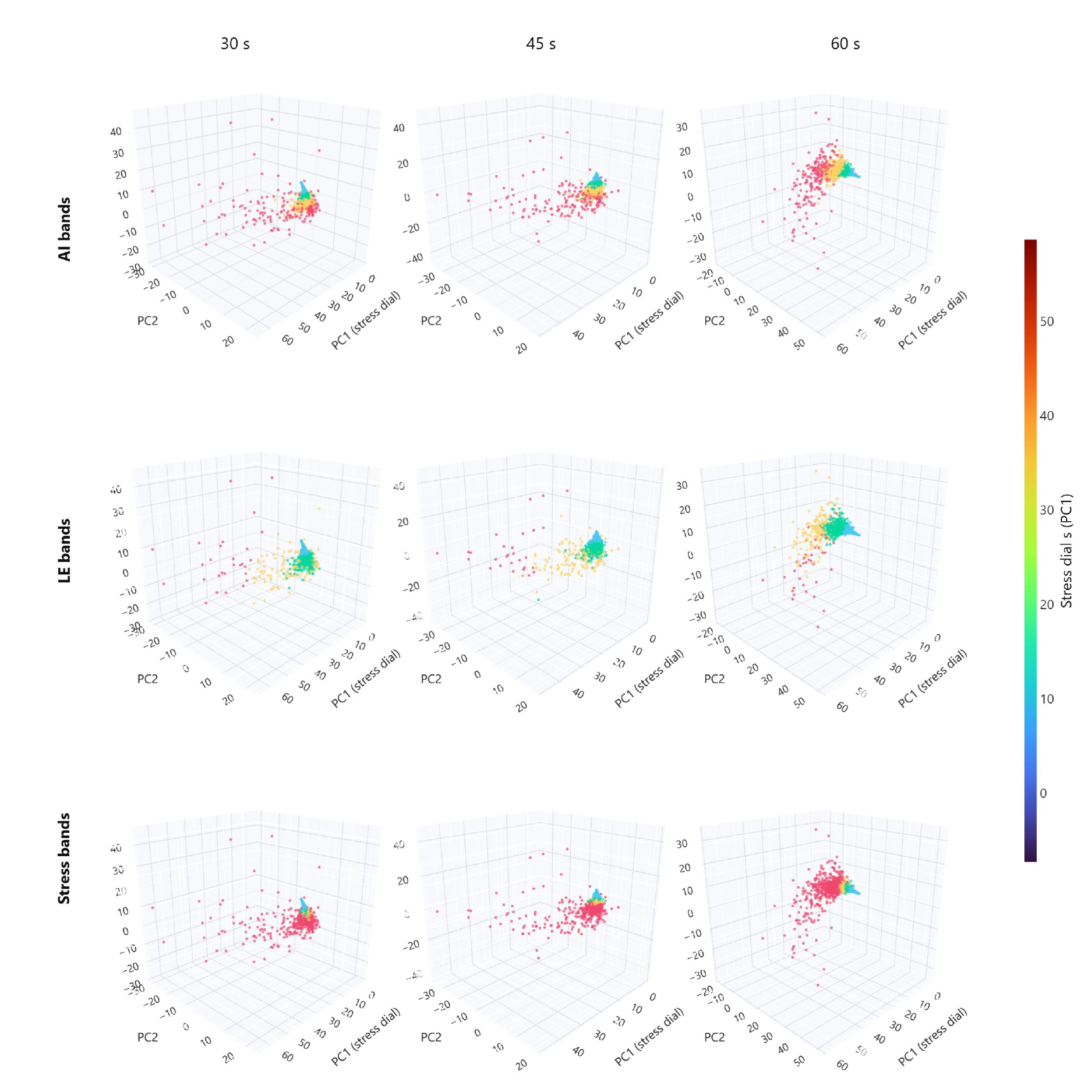}
\caption{Discrete regimes as level-set bands of a continuous stress dial.
Columns show 30, 45 and 60-second covariance grids; rows compare
affine-invariant, Log-Euclidean, and common-axis bands, revealing adjoining
regions rather than separate state clusters.}
\label{fig:hierarchy_bands_3x3}
\end{figure}

\begin{figure}[t]
\centering
\includegraphics[width=0.98\linewidth]{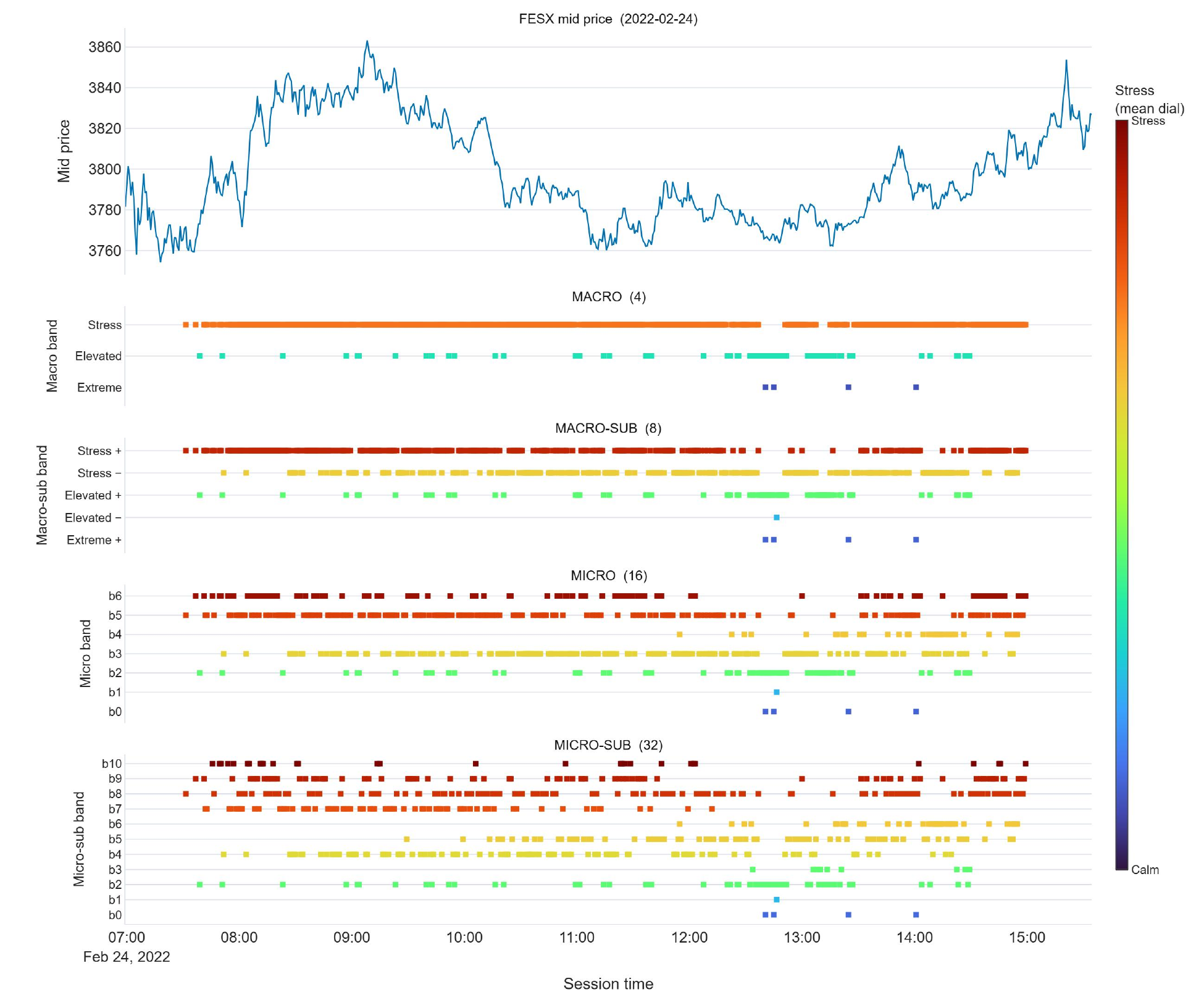}
\caption{A single trading day across the regime hierarchy. The 45-second view
shows how increasingly granular bands partition the same FESX price path
rather than reveal distinct state transitions.}
\label{fig:regime_day_hierarchy}
\end{figure}

Although discrete regimes provide a convenient summary of market conditions,
their value depends on whether the resulting hierarchy transfers out of
sample. We therefore estimate band boundaries using training data only and
evaluate them with chronological and purged walk-forward splits under a
one-day embargo. Figure~\ref{fig:hierarchy_bands_3x3} shows that, across
durations and both SPD metrics, the bands occupy adjoining regions of a
connected cloud and preserve the same stress ordering despite shifts in their
fitted boundaries. Figure~\ref{fig:regime_day_hierarchy} provides the
corresponding intraday view: macro, macro-sub, micro, and micro-sub labels
progressively divide the same price and stress path rather than identify
separate market states. Table~\ref{tab:hierarchy} confirms this interpretation
quantitatively. Moving from macro to micro-sub bands at 30 seconds increases
in-sample variance separation from 0.0861 to 0.2626 under LE and from 0.0610
to 0.2315 under AI, but incremental out-of-sample \(R^2\) falls from
\(+0.0391\) to \(-0.0813\) under LE and from \(+0.0114\) to \(-0.0507\) under
AI. Partition stability also declines, with ARI falling from 0.8974 to 0.6264
under LE and from 0.8799 to 0.4589 under AI. Finer hierarchies therefore
improve in-sample separation at the expense of stability and out-of-sample
transfer, supporting their interpretation as increasingly detailed
subdivisions of a continuum rather than distinct market regimes.

\begin{table}[!t]
\centering
\caption{Hierarchy complexity and transfer at 30 seconds.  ``Single split''
uses the chronological half split; ``pooled WF'' concatenates expanding-year
test losses before block bootstrap.}
\label{tab:hierarchy}
\small
\begin{tabular}{llrrrr}
\toprule
Metric & Tier & Bands & $\eta^2$ single & $\Delta R^2$ single &
$\Delta R^2$ pooled WF\\
\midrule
LE & Macro & 4 & 0.0861 & 0.0391 & 0.02919\\
LE & Macro-sub & 8 & 0.2047 & $-0.0432$ & 0.01157\\
LE & Micro & 16 & 0.2384 & $-0.0837$ & 0.00929\\
LE & Micro-sub & 27 & 0.2626 & $-0.0813$ & $-0.00124$\\
\addlinespace
AI & Macro & 4 & 0.0610 & 0.0114 & 0.04315\\
AI & Macro-sub & 8 & 0.1570 & $-0.0310$ & 0.01835\\
AI & Micro & 16 & 0.1715 & $-0.0414$ & 0.01741\\
AI & Micro-sub & 31 & 0.2315 & $-0.0507$ & 0.01022\\
\bottomrule
\end{tabular}
\begin{minipage}{0.94\linewidth}\footnotesize
Positive pooled AI micro-sub performance does not supersede the adverse 2025
fold.
\end{minipage}
\end{table}

\begin{figure}[t]
\centering
\includegraphics[width=0.98\linewidth]{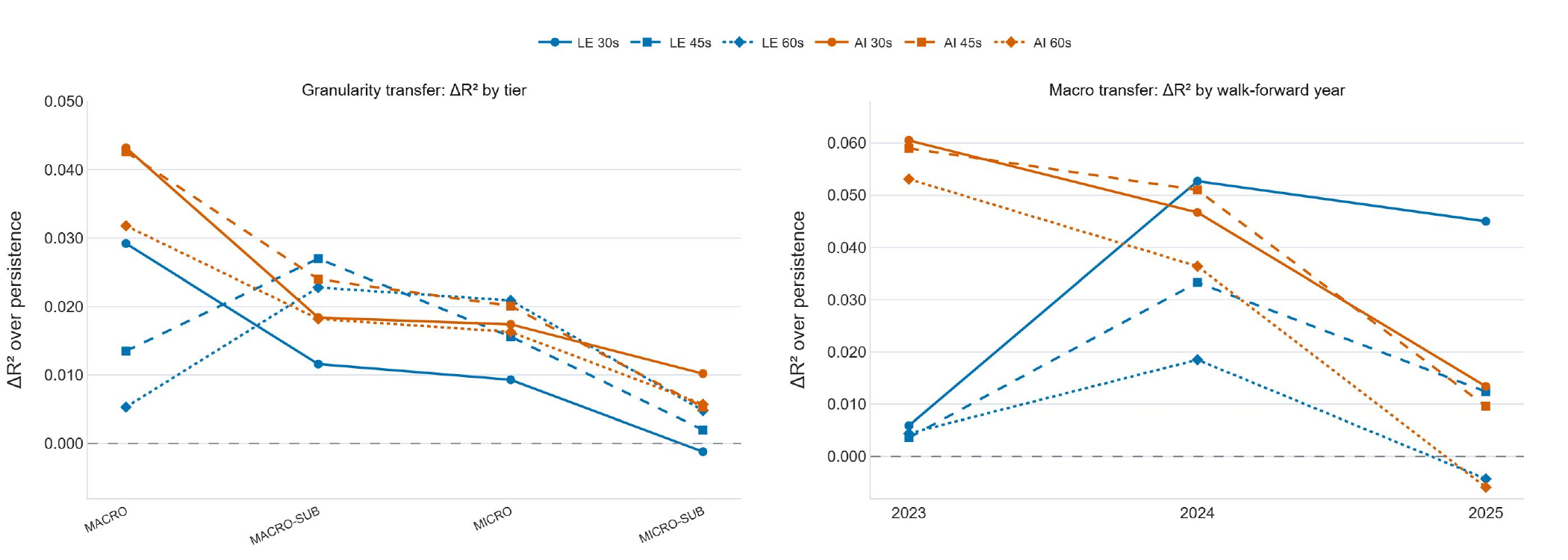}
\caption{Hierarchy transfer weakens with granularity and time. Purged
walk-forward incremental OOS $R^2$ relative to persistence with a one-day
embargo, shown by hierarchy tier (left) and macro-band fold and duration
(right).}
\label{fig:hierarchy_transfer_2panel}
\end{figure}

Figure~\ref{fig:hierarchy_transfer_2panel} verifies that the hierarchy-overfit
pattern extends beyond a single train-test split. Across expanding-year folds,
the pooled LE gain declines from \(+0.02919\) for macro bands to
\(-0.00124\) for micro-sub bands, whose confidence interval
\([-0.00692, 0.00469]\) includes zero. Although pooled AI estimates remain
positive, the advantage weakens as market conditions change; by 2025, the AI
macro estimate falls to \(+0.01338\) with an interval that includes zero, and
all finer tiers become negative. Only 30-second LE retains a clear macro-band
gain in 2025. These results give discrete regimes a limited but useful role:
coarse bands can add predictive information beyond persistence, whereas finer
hierarchies primarily capture historical density and fail to transfer out of
sample. The continuous stress dial therefore remains the primary empirical
object, with discrete bands serving as coarse reporting categories and the
supporting evaluation and frozen-transfer diagnostics reported in the
supplementary materials.

\FloatBarrier
\subsection{Continuous Stress Dynamics and Online Monitoring}
\label{sec:results:dial}\linkdest{sec:results:dial}

\begin{figure}[t]
\centering
\includegraphics[width=0.98\linewidth]{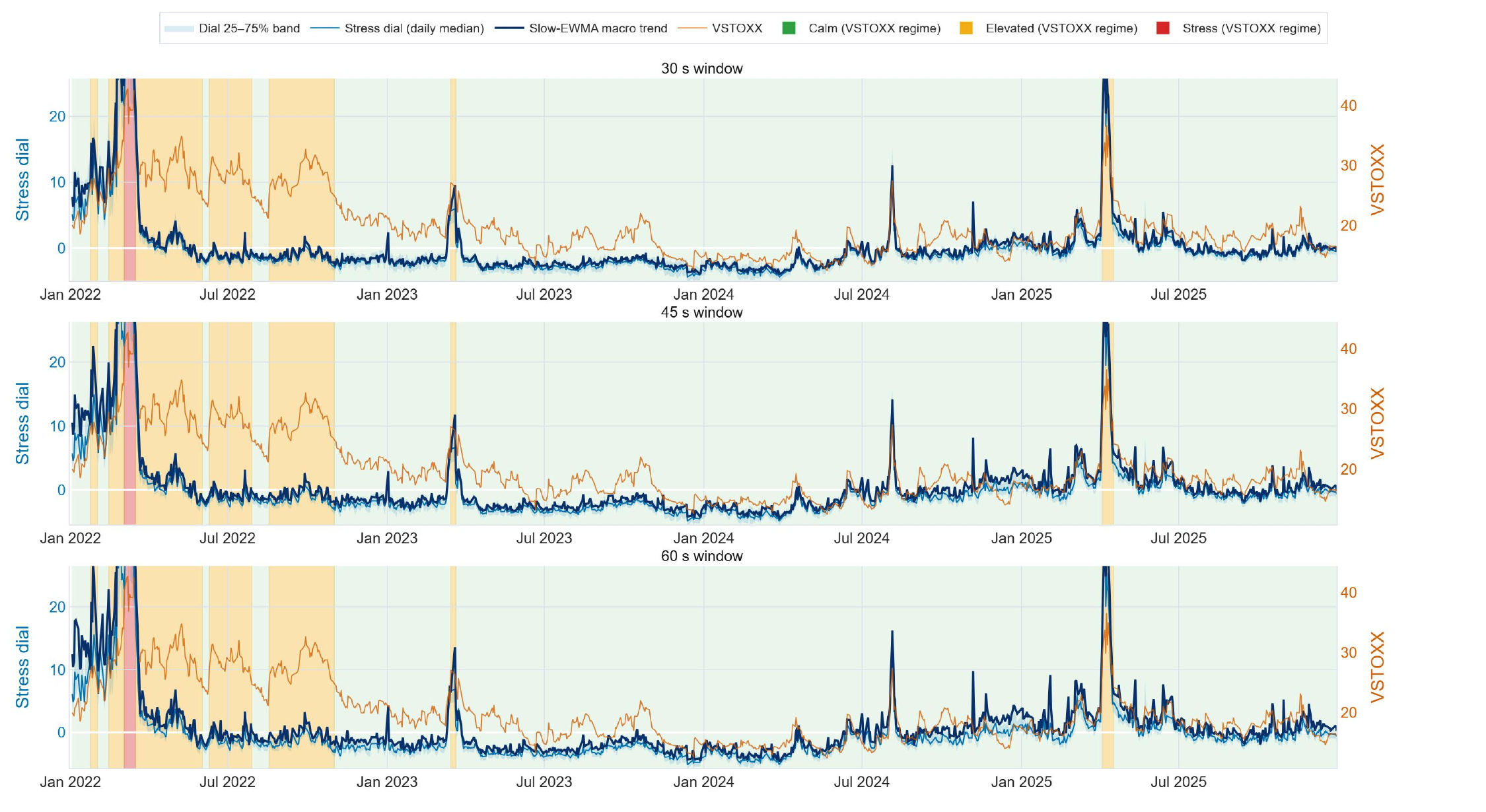}
\caption{Continuous stress-dial dynamics across covariance durations,
2022-2025.}
\label{fig:stress_dial_continuous}
\end{figure}

\begin{figure}[t]
\centering
\includegraphics[width=0.98\linewidth]{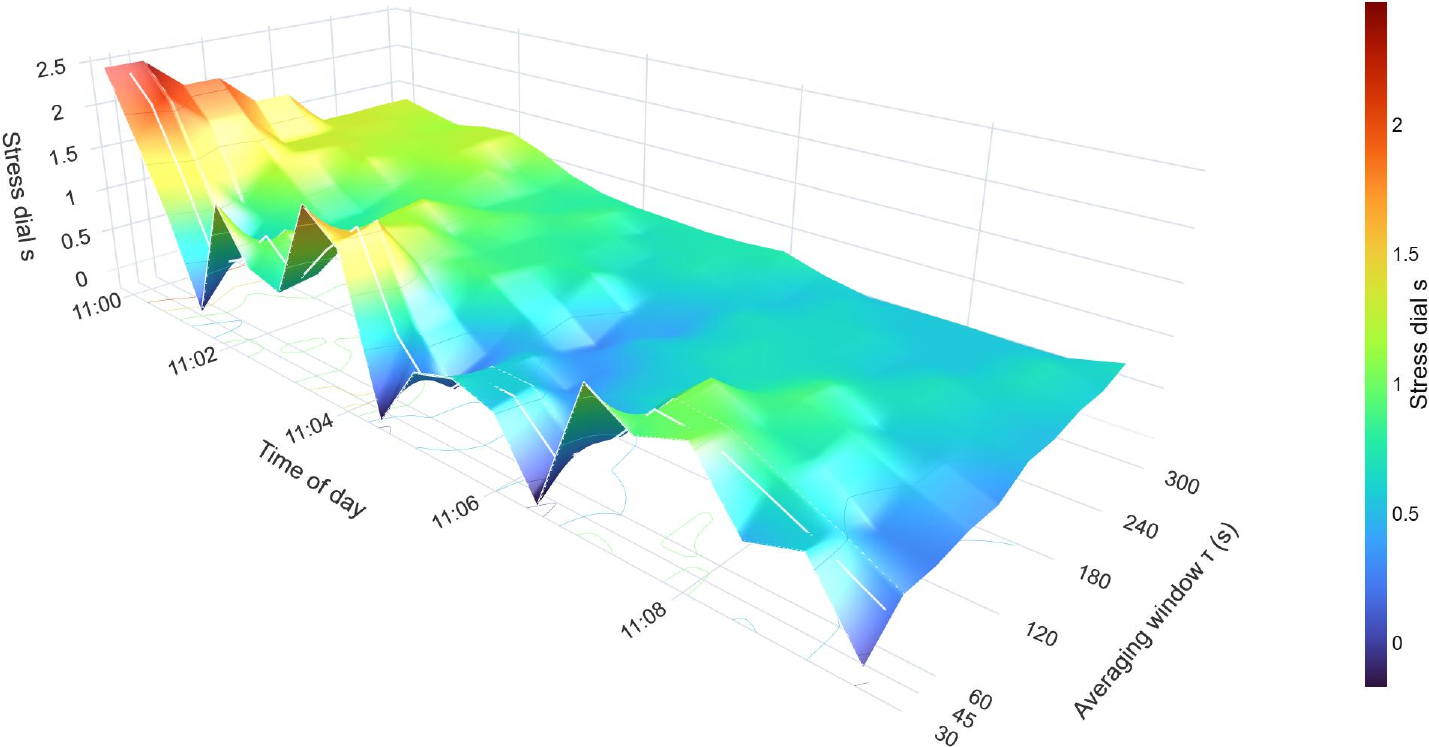}
\caption{Continuous order-book state across observation windows.}
\label{fig:lob_continuity_surface}
\end{figure}

A credible continuous state representation should remain stable as the
observation window changes. Figures~\ref{fig:stress_dial_continuous}
and~\ref{fig:lob_continuity_surface} provide this test: the 30, 45 and
60-second estimates preserve the same timing and economic ordering, with
longer windows producing smoother paths, while the finer-resolution surface
shows a gradual transition as the averaging window increases from 30 to 300
seconds. Table~\ref{tab:continuous_state_prediction} confirms that the
resulting dial is also economically informative. Lag-one autocorrelation rises
from 0.9391 at 30 seconds to 0.9570 at 60 seconds, while the VSTOXX correlation
increases from 0.4997 to 0.5221. The slow dial adds statistically positive
out-of-sample predictive content beyond persistence at every duration,
although its incremental \(R^2\) declines from 0.0328 to 0.0188 as the window
lengthens; the fast residual contributes a smaller gain throughout. Longer
windows therefore strengthen persistence and external alignment while leaving
less short-horizon information for the slow component to add. The
corresponding one-window predictive-coincidence and directional-accuracy
diagnostics are reported in the supplementary materials.

\begin{table}[t]
\centering
\caption{Continuous-state validity and selected-timescale predictive performance.}
\label{tab:continuous_state_prediction}
\scriptsize
\setlength{\tabcolsep}{2.5pt}
\textit{Panel A: State validity and persistence}\\[0.5ex]
\begin{tabular}{lrrrrrr}
\toprule
Duration & Windows & \shortstack{PC1\\fraction} &
\shortstack{Corr.\\VSTOXX} & \shortstack{Pearson\\next RV} &
\shortstack{Spearman\\next RV} & \shortstack{Lag-1 AC /\\decay HL}\\
\midrule
30 s & 837,890 & 0.2741 & 0.4997 & 0.3828 & 0.2955 & 0.9391 / 11.03\\
45 s & 559,261 & 0.2855 & 0.5141 & 0.4145 & 0.3439 & 0.9515 / 13.93\\
60 s & 419,493 & 0.2922 & 0.5221 & 0.4332 & 0.3775 & 0.9570 / 15.78\\
\bottomrule
\end{tabular}

\vspace{1.2ex}
\textit{Panel B: Five-window-half-life predictive performance}\\[0.5ex]
\begin{tabular}{lrrrr}
\toprule
Duration & \shortstack{Best\\$h$} & \shortstack{Persist / +slow / +fast\\$R^2$} &
\shortstack{Slow $\Delta R^2$\\(95\% interval)} &
\shortstack{Fast\\$\Delta R^2$}\\
\midrule
30 s & 5 & 0.4273 / 0.4601 / 0.4694 & 0.0328 [0.0282, 0.0378] & 0.0092\\
45 s & 5 & 0.5225 / 0.5470 / 0.5535 & 0.0244 [0.0213, 0.0278] & 0.0066\\
60 s & 5 & 0.5871 / 0.6060 / 0.6103 & 0.0188 [0.0165, 0.0213] & 0.0044\\
\bottomrule
\end{tabular}
\begin{minipage}{0.94\linewidth}\footnotesize
M0 is the persistence model, M1 adds the slow stress dial, and M2 adds the
fast residual and velocity. Predictive results are pooled purged
walk-forward estimates with a one-day embargo; standardisation, the PCA axis,
and its VSTOXX orientation are refitted on each training fold. \(h\) is measured in windows.
\end{minipage}
\end{table}

The dial also provides an economically meaningful ordering of future
volatility. Mean next-window realised volatility rises monotonically across
dial deciles at every duration, while the highest decile contains 5.32 to
5.72 times the unconditional incidence of the one-percent volatility tail.
The dial therefore captures both gradual changes in market conditions and
concentrations of extreme short-horizon risk. Its dynamics are persistent but
mean-reverting: the pooled Ornstein-Uhlenbeck estimates are positive at every
duration, with confidence intervals excluding zero and implied half-lives of
approximately 11 to 15 windows. An independent transfer-operator analysis
produces the same ordering across durations, and the complete dynamics
certificate passes in every case, as reported in the supplementary materials.
The dial's right-skewed distribution is consistent with long calm periods
interrupted by brief stress episodes, although these estimates describe local
empirical relaxation rather than a globally Gaussian, stationary, or Markov
process.

\begin{figure}[t]
\centering
\includegraphics[width=0.98\linewidth]{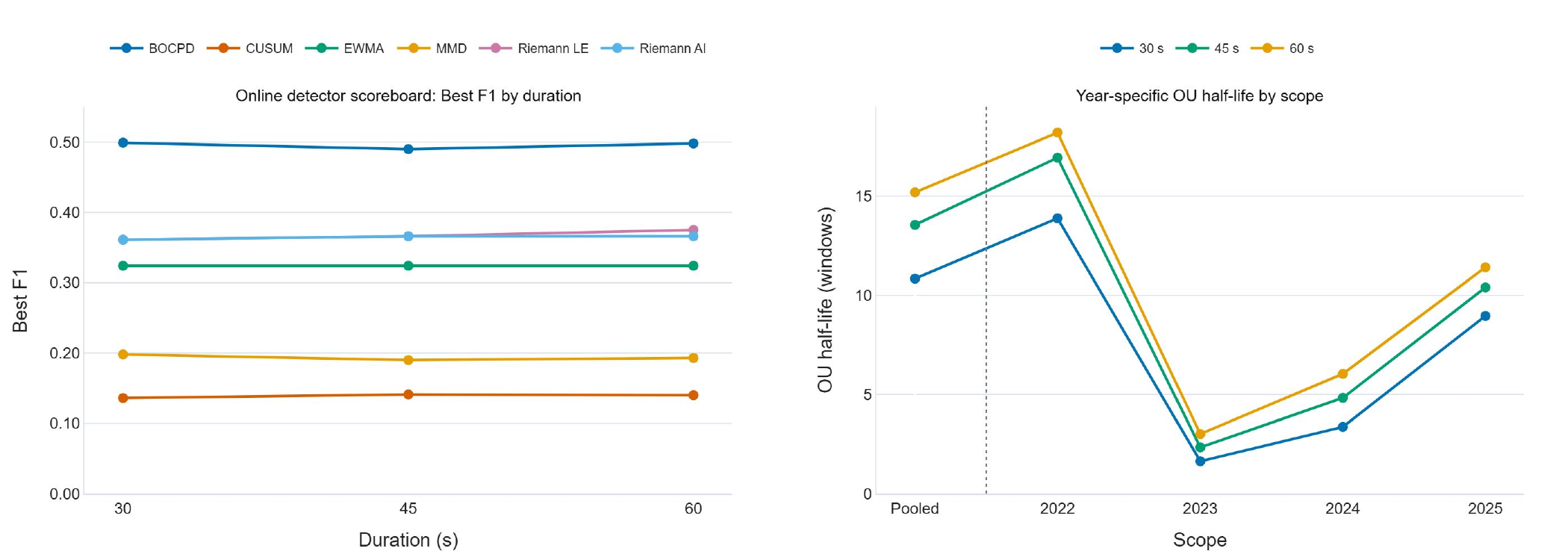}
\caption{Online detector performance and year-specific stress relaxation.}
\label{fig:detector_ou_2panel}
\end{figure}

Figure~\ref{fig:detector_ou_2panel} shows that mean reversion varies over
time, with faster relaxation in 2023 and slower relaxation in 2022, while the
ordering across covariance durations remains stable. Its left panel evaluates
whether the continuous dial can also support online monitoring. BOCPD
achieves the highest \(F_1\) score at every duration, approximately 0.49 to
0.50, followed by the Riemannian detectors. EWMA is precise but detects
relatively few events, with precision 0.857 and recall 0.20, while CUSUM and
MMD perform less strongly. The supplementary materials confirm that the LE
and AI Riemannian implementations produce nearly identical \(F_1\), recall,
and AUC-PR values, showing that detector performance is robust to the choice
of SPD metric, although geometry alone does not solve the event-detection
problem. Annual results provide an important qualification: the
dial-VSTOXX correlation strengthens from approximately 0.42 in 2022 to
0.75-0.77 in 2025, BOCPD \(F_1\) reaches 0.745 in 2025, and Riemannian LE
performance peaks in 2024. External validity and detector performance
therefore vary over time but remain consistent across covariance durations.
The dial is consequently best used within a layered monitoring system
alongside a separate change alarm rather than as an automatic crisis
declaration.

\FloatBarrier
\begin{figure}[t]
\centering
\includegraphics[width=0.98\linewidth]{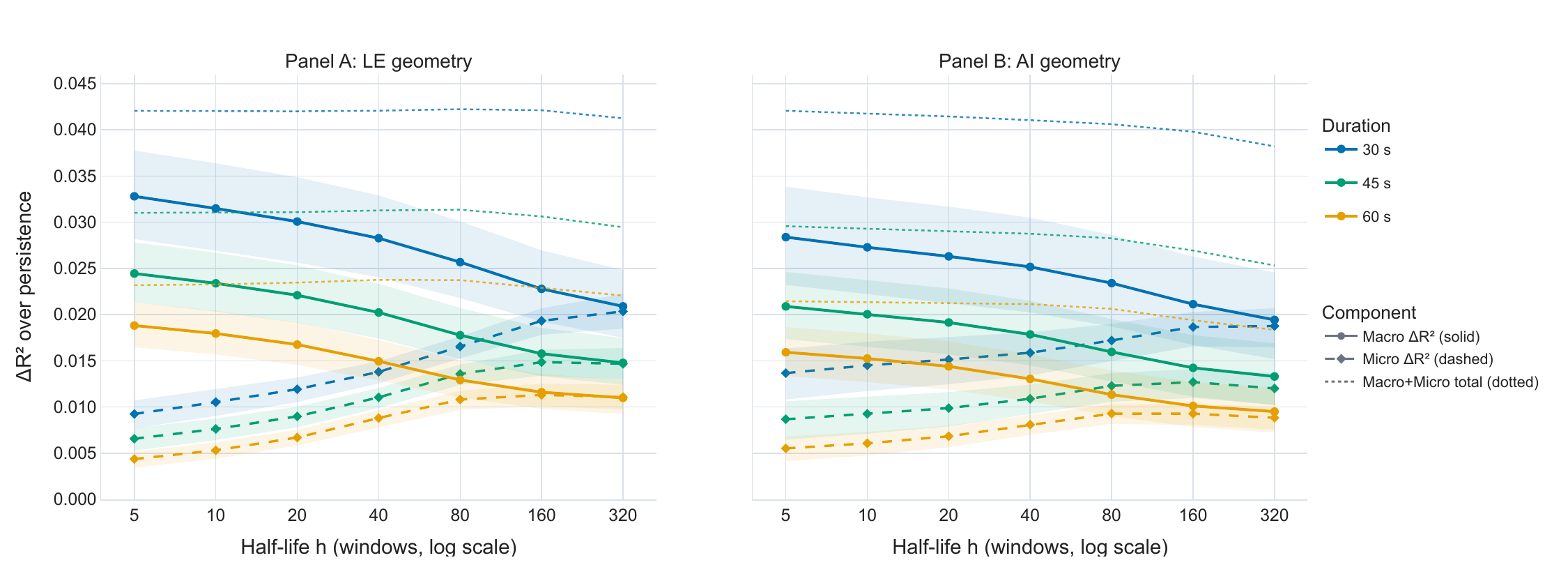}
\caption{Slow-fast predictive contributions under LE (Panel A) and AI (Panel B).}
\label{fig:granularity_sweep}
\end{figure}

\begin{table}[t]
\centering
\caption{Continuous micro-subspace model ladder under LE and AI.}
\label{tab:micro_subspace_pooled}
\scriptsize
\setlength{\tabcolsep}{2.4pt}
\textit{Panel A: LE geometry}\\[0.5ex]
\begin{tabular}{lrrrrrrr}
\toprule
Duration & \shortstack{M0\\\(R^2\)} & \shortstack{M1\\\(R^2\)} &
\shortstack{M2\\\(R^2\)} & \shortstack{Macro\\\(\Delta R^2\)} &
\shortstack{Micro\\\(\Delta R^2\)} &
\shortstack{Micro 95\%\\interval} & Significant\\
\midrule
30 s & \(-0.0886\) & 0.0123 & 0.0340 & 0.1010 & 0.0216 & [0.0118, 0.0299] & Yes\\
45 s & \(-0.0754\) & 0.0380 & 0.0648 & 0.1134 & 0.0268 & [0.0175, 0.0354] & Yes\\
60 s & \(-0.0672\) & 0.0541 & 0.0855 & 0.1213 & 0.0315 & [0.0211, 0.0412] & Yes\\
\bottomrule
\end{tabular}

\vspace{1.2ex}
\textit{Panel B: AI geometry}\\[0.5ex]
\begin{tabular}{lrrrrrrr}
\toprule
Duration & \shortstack{M0\\\(R^2\)} & \shortstack{M1\\\(R^2\)} &
\shortstack{M2\\\(R^2\)} & \shortstack{Macro\\\(\Delta R^2\)} &
\shortstack{Micro\\\(\Delta R^2\)} &
\shortstack{Micro 95\%\\interval} & Significant\\
\midrule
30 s & \(-0.0886\) & 0.0123 & 0.0339 & 0.1010 & 0.0215 & [0.0117, 0.0299] & Yes\\
45 s & \(-0.0754\) & 0.0380 & 0.0645 & 0.1134 & 0.0265 & [0.0172, 0.0351] & Yes\\
60 s & \(-0.0672\) & 0.0541 & 0.0851 & 0.1213 & 0.0310 & [0.0207, 0.0408] & Yes\\
\bottomrule
\end{tabular}
\begin{minipage}{0.94\linewidth}\footnotesize
M0 is persistence, M1 adds the macro dial, and M2 adds the first continuous
principal component of the within-day residual tangent space. Panels A and B
report pooled out-of-sample estimates under LE and AI, respectively. The micro
increment is positive and statistically significant for both geometries.
\end{minipage}
\end{table}

\begin{figure}[t]
\centering
\includegraphics[width=0.82\linewidth]{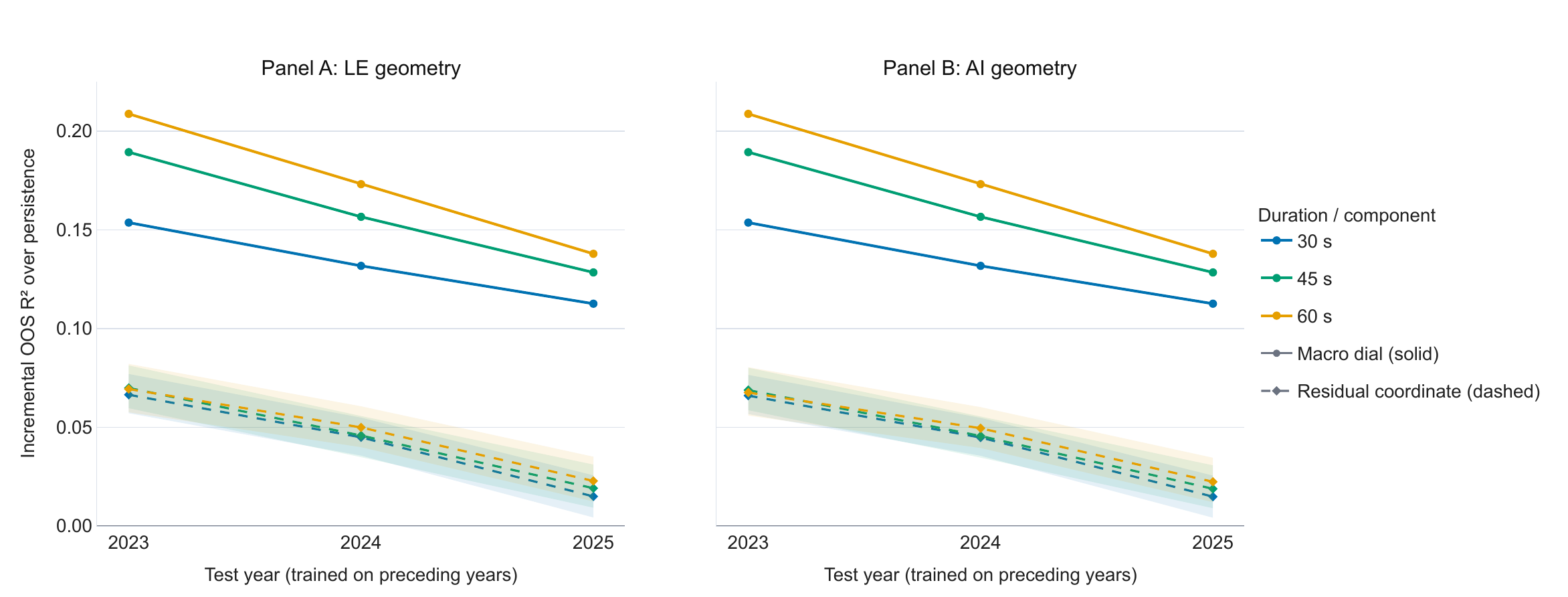}
\caption{Forward validation of continuous residual-subspace signals under LE
(Panel A) and AI (Panel B). Solid lines report the macro contribution; dashed
lines and bands report the residual-coordinate increment and its 95\%
day-block confidence interval.}
\label{fig:micro_subspace_forward}
\end{figure}

Figure~\ref{fig:granularity_sweep} extends the continuum result to the
timescale dimension, using dial coordinates estimated train-only within each
fold. The same profile appears under LE and AI: as the EWMA
half-life increases, the macro contribution declines while the micro
contribution rises, and their combined incremental \(R^2\) changes only
modestly. The macro-micro split therefore represents a smooth decomposition
of the same process rather than a distinct timescale boundary.
Table~\ref{tab:micro_subspace_pooled} and
Figure~\ref{fig:micro_subspace_forward} then test whether predictive structure
remains after accounting for the macro dial. The model ladder compares
persistence (M0), the addition of the macro dial (M1), and the addition of a
continuous residual coordinate (M2). The macro dial provides the dominant
incremental \(R^2\), ranging from 0.101 to 0.121 across durations, while the
residual coordinate adds a smaller but statistically significant 0.022 to
0.031. Forward validation preserves this ordering across all duration-year
cells, with residual confidence intervals remaining positive but weakening
toward 2025. These findings clarify the effective dimension of
approximately 1.4: one continuous stress axis dominates, but modest
predictive structure remains in the residual space. Importantly, this
additional signal is captured by a continuous coordinate rather than the
discrete subdivisions that fail out of sample in
Section~\linkref{sec:results:hierarchy}. Together, these results describe a
continuous, economically ordered, and mean-reverting stress state. Its
predictive content varies smoothly across timescales, remains robust across
durations and SPD metrics, and can support online monitoring, although
detection quality still depends on the algorithm, threshold, and evaluation
year.

\FloatBarrier

\subsection{Intraday Volatility Forecasting}\label{sec:results:intraday}

\begin{figure}[!b]
\centering
\includegraphics[width=0.98\linewidth]{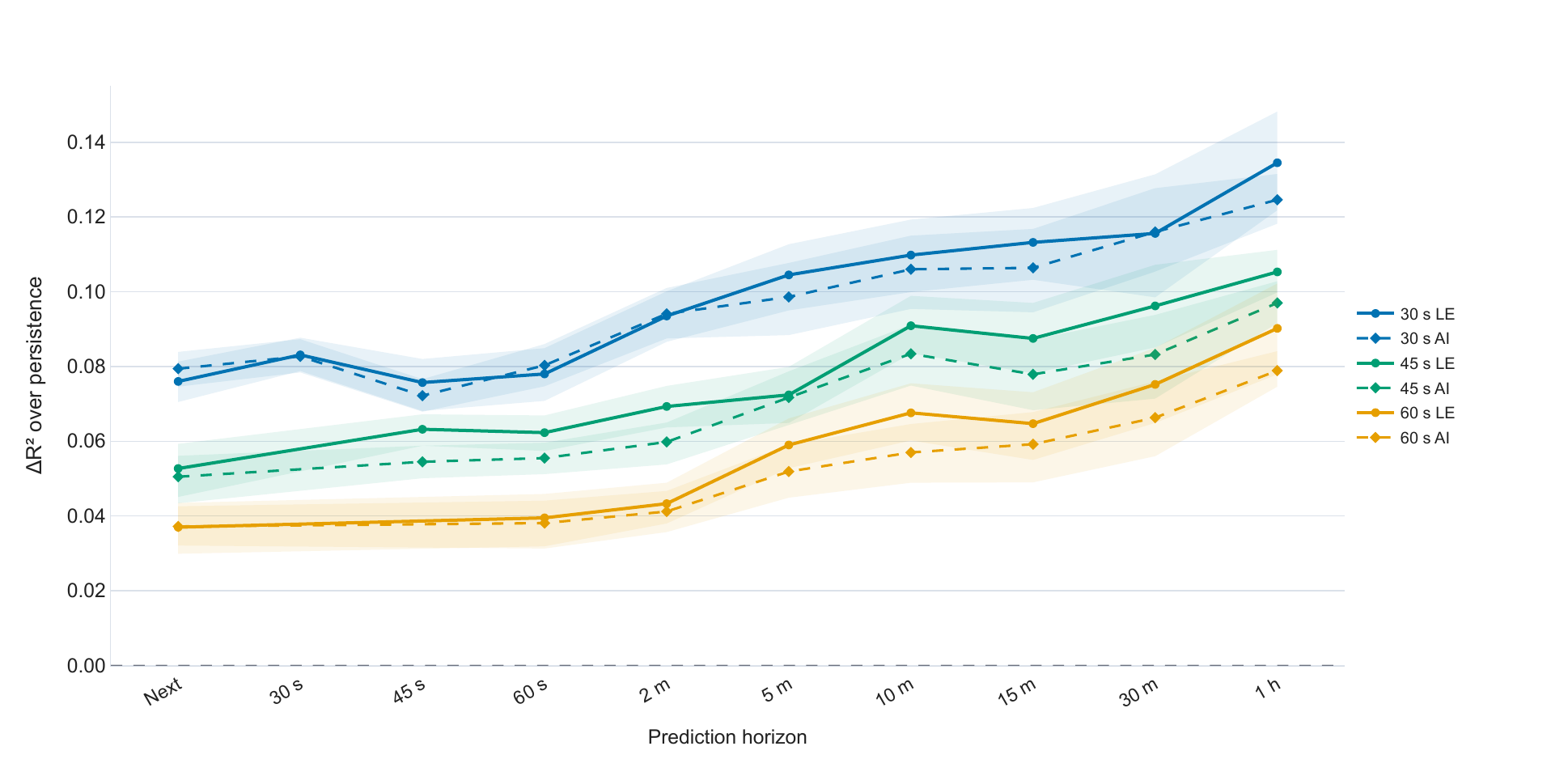}
\caption{Intraday incremental \(R^2\) across durations and horizons.}
\label{fig:incremental_r2_grid}
\end{figure}

The intraday experiment contains 54 feasible duration-horizon-metric cells.
Six combinations are excluded mechanically because the forecast horizon is
shorter than the covariance window.  In every feasible cell, the best model
produces a positive incremental out-of-sample \(R^2\), with its 95\% day-block
confidence interval bounded above zero.  Figure~\ref{fig:incremental_r2_grid}
summarises this central result across all horizons, covariance durations, and
geometries.  The predictive advantage generally increases with the forecast
horizon and remains positive across all covariance durations, indicating that
the covariance state becomes relatively more informative at longer horizons.

Absolute and incremental performance peak at different horizons.
Table~\ref{tab:forecast_map} shows that the ten-minute LE-SVR forecast raises
out-of-sample \(R^2\) above persistence from 0.5726 to 0.6821 at 30 seconds,
from 0.6370 to 0.7273 at 45 seconds, and from 0.6821 to 0.7483 at 60 seconds.
Absolute \(R^2\) subsequently declines, but the relative advantage continues
to widen because persistence deteriorates more quickly.  The covariance state
therefore provides its largest relative benefit at longer horizons, although
total forecast fit is highest near ten minutes.  Across covariance durations,
the same positive horizon profile appears under both LE and AI, showing that
the result is not driven by a particular geometry or temporal resolution.
Even the linear model consistently outperforms persistence across the feasible
grid.

\begin{table}[!b]
\centering
\caption{Selected LE-SVR intraday forecast results.  $\Delta R^2$ is relative
to the cell-specific persistence forecast.}
\label{tab:forecast_map}
\small
\begin{tabular}{lrrrrrr}
\toprule
 & \multicolumn{2}{c}{30 s covariance} &
\multicolumn{2}{c}{45 s covariance} &
\multicolumn{2}{c}{60 s covariance}\\
\cmidrule(lr){2-3}\cmidrule(lr){4-5}\cmidrule(lr){6-7}
Horizon & $R^2$ & $\Delta R^2$ & $R^2$ & $\Delta R^2$ & $R^2$ & $\Delta R^2$\\
\midrule
2 min & 0.6295 & 0.0938 & 0.6354 & 0.0644 & 0.6758 & 0.0451\\
5 min & 0.6820 & 0.1104 & 0.7017 & 0.0714 & 0.7335 & 0.0578\\
10 min & 0.6821 & 0.1096 & 0.7273 & 0.0903 & 0.7483 & 0.0662\\
30 min & 0.6457 & 0.1117 & 0.6914 & 0.0928 & 0.7173 & 0.0717\\
1 hour & 0.6162 & 0.1313 & 0.6488 & 0.1000 & 0.6821 & 0.0865\\
\bottomrule
\end{tabular}
\begin{minipage}{0.94\linewidth}\footnotesize
All displayed increments have block-bootstrap intervals above zero.
\end{minipage}
\end{table}

\begin{figure}[t]
\centering
\includegraphics[width=0.98\linewidth]{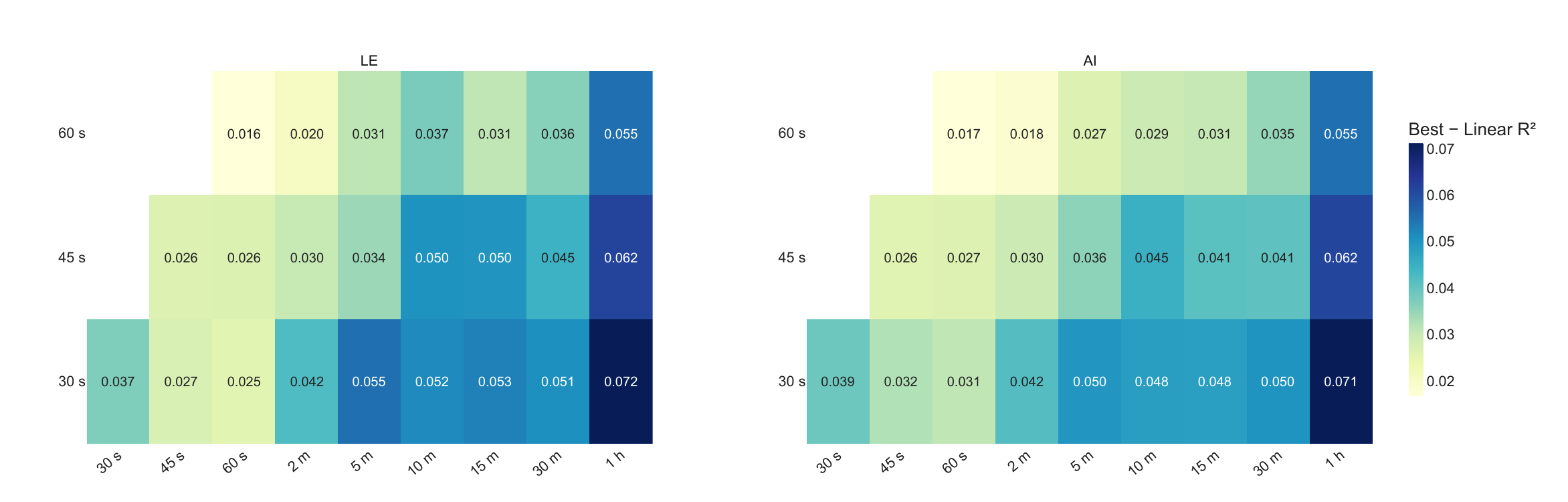}
\caption{Nonlinear capacity gains across horizons and covariance durations.}
\label{fig:capacity_gain}
\end{figure}

The dial-to-volatility relationship is only mildly nonlinear.
Figure~\ref{fig:capacity_gain} reports the gain in out-of-sample \(R^2\) from
replacing the linear model with the best learner.  These gains are positive
but modest, ranging from 0.015 to 0.072.  At 30 seconds, the capacity gain is
0.052 at ten minutes and 0.072 at one hour.  A linear model therefore captures
most absolute forecast fit, while nonlinear estimation enlarges the
incremental advantage over persistence.

\begin{figure}[t]
\centering
\includegraphics[width=0.92\linewidth]{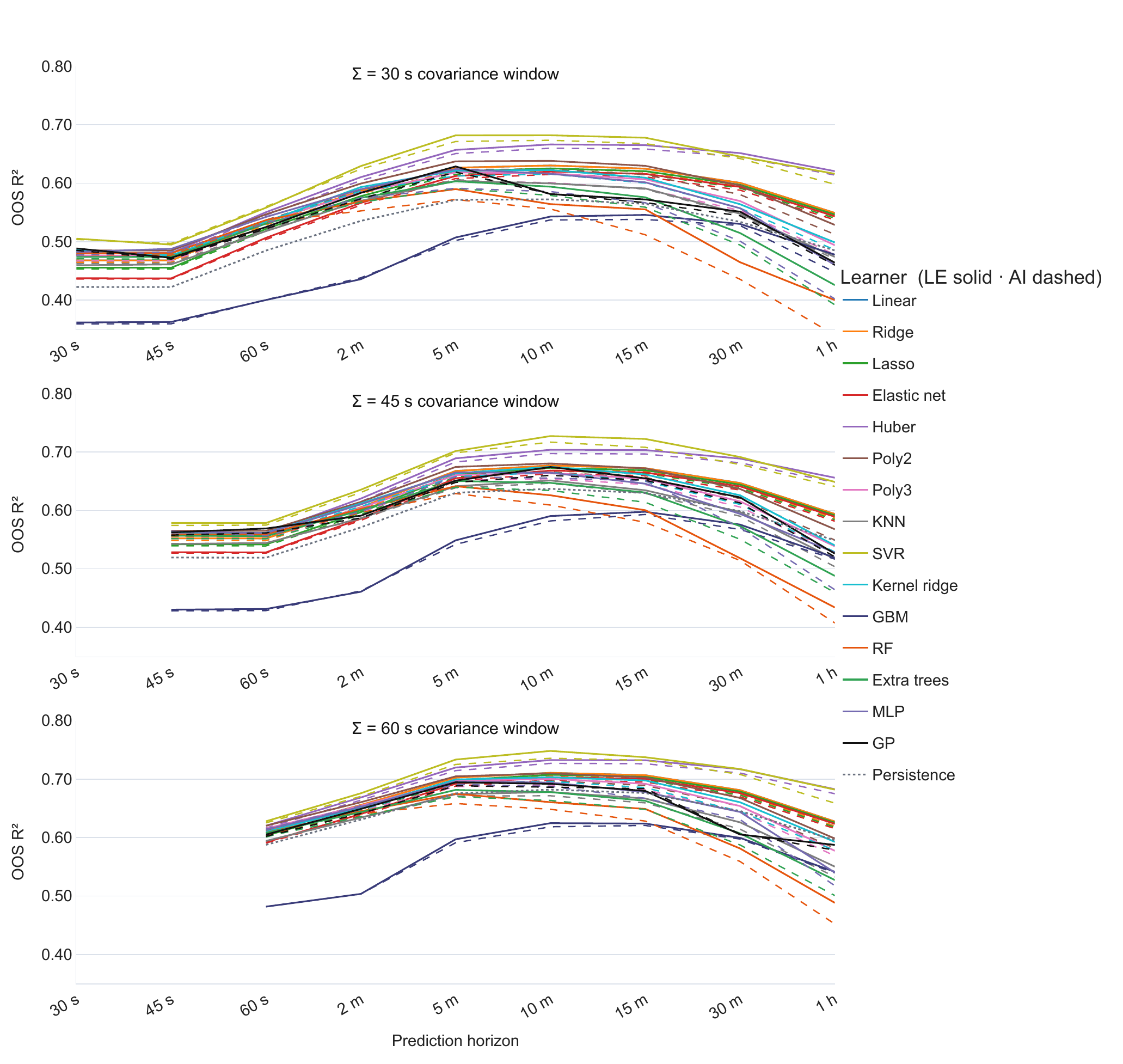}
\caption{Forecast performance across learners, horizons, and covariance
durations.}
\label{fig:oos_r2_overlay_stack}
\end{figure}

The result is also robust across forecasting architectures.
Figure~\ref{fig:oos_r2_overlay_stack} compares all fifteen learners and shows
a common hump-shaped absolute-\(R^2\) profile that peaks near the ten-minute
horizon across covariance durations and SPD metrics.  Support-vector
regression leads in 44 of the 54 cells, with
Huber regression leading in the remaining ten.  Nevertheless, mean
out-of-sample \(R^2\) is 0.605 for the linear model and 0.641 for SVR.  The
stable horizon profile across learners is therefore more important than the
identity of the winning estimator.

Metric and loss comparisons provide additional robustness.  LE and AI produce
closely aligned horizon profiles, although individual incremental-\(R^2\)
differences can reach approximately 0.01.  At the 30-second ten-minute
horizon, for example, the increments are 0.1098 under LE and 0.1060 under AI.
Replacing \(R^2\) with QLIKE preserves the same broad horizon and learner
ordering.  The forecast result is therefore robust, but the two geometries
should not be treated as numerically identical.

Directional accuracy supplies complementary next-window evidence.  The
continuous predictive model correctly anticipates the direction of the next
realised-volatility change in 61.2\%, 60.5\%, and 59.5\% of observations at
the 30, 45 and 60-second covariance windows.  Accuracy declines modestly at
coarser resolutions, consistent with their smaller short-horizon incremental
\(R^2\).  Because these estimates come from the continuous predictive block
rather than the complete Step~7 grid, they are supporting evidence rather
than a separate forecast-map certificate.

Table~\ref{tab:year_specific_intraday} provides the principal temporal
qualification.  At the 30-second ten-minute LE cell, incremental \(R^2\)
declines from 0.1721 in 2023 to 0.0753 in 2024 and 0.0325 in 2025.  Similar
weakening appears at coarser windows, with the 60-second ten-minute LE
increment reaching 0.0044 in 2025.  One-hour performance is more persistent
but remains heterogeneous across metrics and years.

\begin{table}[t]
\centering
\caption{Year-specific intraday out-of-sample \(R^2\) and incremental \(R^2\).}
\label{tab:year_specific_intraday}
\scriptsize
\setlength{\tabcolsep}{5pt}
\begin{tabular}{llrrrr}
\toprule
Duration & \shortstack{Horizon and\\metric/model} &
2023 & 2024 & 2025\\
\midrule
30 s & 10 min, LE/SVR & 0.7814 / 0.1721 & 0.5893 / 0.0753 & 0.5914 / 0.0325\\
30 s & 10 min, AI/SVR & 0.7669 / 0.1576 & 0.5598 / 0.0459 & 0.6102 / 0.0512\\
45 s & 10 min, LE/SVR & 0.8119 / 0.1421 & 0.6532 / 0.0679 & 0.6460 / 0.0216\\
45 s & 10 min, AI/SVR & 0.8011 / 0.1313 & 0.6249 / 0.0396 & 0.6512 / 0.0268\\
60 s & 10 min, LE/SVR & 0.8329 / 0.1201 & 0.6703 / 0.0315 & 0.6709 / 0.0044\\
60 s & 10 min, AI/SVR & 0.8211 / 0.1084 & 0.6393 / 0.0004 & 0.6722 / 0.0058\\
\midrule
30 s & 1 hour, LE/Huber & 0.6358 / 0.1284 & 0.6081 / 0.1687 & 0.6054 / 0.1217\\
30 s & 1 hour, AI/Huber & 0.6221 / 0.1147 & 0.6026 / 0.1632 & 0.6095 / 0.1259\\
45 s & 1 hour, LE/Huber & 0.6708 / 0.1027 & 0.6450 / 0.1345 & 0.6397 / 0.0928\\
45 s & 1 hour, AI/Huber & 0.6540 / 0.0859 & 0.6411 / 0.1306 & 0.6433 / 0.0964\\
60 s & 1 hour, LE/Huber & 0.6988 / 0.0852 & 0.6736 / 0.1094 & 0.6632 / 0.0728\\
60 s & 1 hour, AI/Huber & 0.6817 / 0.0681 & 0.6714 / 0.1072 & 0.6667 / 0.0763\\
\bottomrule
\end{tabular}
\begin{minipage}{0.94\linewidth}\footnotesize
Each annual entry reports test \(R^2\) followed by incremental \(R^2\) over
the corresponding persistence benchmark.
\end{minipage}
\end{table}

Overall, the covariance stress dial contains statistically and economically
meaningful information for intraday volatility forecasting.  The result
survives changes in horizon, covariance duration, SPD metric, learner, and
loss function.  Its relative advantage is greatest at longer horizons, its
absolute fit peaks near ten minutes, and its annual contribution weakens
toward 2025.
Any trading application would additionally require latency, slippage,
transaction-cost, and capacity evaluation.

\subsection{Next-Day Volatility Forecasting}
\label{sec:results:nextday}

At the daily horizon, 964 of 987 days have complete predictors, with 230
training days and 734 observations in the 2023-2025 test folds.  HAR achieves
OOS \(R^2=0.8298\) and QLIKE 0.0802.  Adding LOB information modestly improves
performance to \(R^2=0.8335\) and QLIKE 0.0760, with a Diebold-Mariano
\(p\)-value of 0.0284.  HAR-LOB is the only model retained by the 90\%
model-confidence set, supporting a small LOB refinement rather than
covariance-regime success.

Adding covariance representation is not robust.  LE reaches \(R^2=0.8374\),
but QLIKE worsens to 0.0804 and the Diebold-Mariano \(p\)-value against HAR is
0.927.  AI lowers \(R^2\) to 0.8197 and raises QLIKE to 0.1099.
Table~\ref{tab:nextday_validation} shows that permutation importance is
slightly positive for LE but negative for AI, and that all Mincer-Zarnowitz
joint calibration tests are rejected.  The LE absolute \(R^2\) increase over
HAR-LOB is 0.0039 and does not survive across metrics, losses, or learners.

The ablation and 15-learner comparison show that covariance representation
alone is a weak daily predictor, while regularised linear learners dominate
the full specification.  The conclusion is horizon-specific: covariance
information is meaningful intraday but does not robustly improve daily HAR
forecasts, consistent with the short mean-reversion timescale documented in
Section~\linkref{sec:results:dial}.

\begin{table}[t]
\centering
\caption{Next-day permutation and Mincer-Zarnowitz validation summary.}
\label{tab:nextday_validation}
\scriptsize
\setlength{\tabcolsep}{3.2pt}
\begin{tabular}{llrrrrrrr}
\toprule
Duration & Metric & \shortstack{Full\\\(R^2\)} &
\shortstack{Permuted\\\(R^2\)} & Drop & Survives &
\shortstack{MZ\\\(a\)} & \shortstack{MZ\\\(b\)} &
\shortstack{Joint\\\(p\)}\\
\midrule
30 s & LE & 0.8374 & 0.8288 & \(+0.0087\) & Yes & 0.00129 & 0.8866 & \(<0.001\)\\
30 s & AI & 0.8197 & 0.8280 & \(-0.0083\) & No  & 0.00205 & 0.8861 & \(<0.001\)\\
45 s & LE & 0.8375 & 0.8307 & \(+0.0068\) & Yes & 0.00135 & 0.8868 & \(<0.001\)\\
45 s & AI & 0.8191 & 0.8297 & \(-0.0106\) & No  & 0.00215 & 0.8777 & \(<0.001\)\\
60 s & LE & 0.8342 & 0.8305 & \(+0.0037\) & Yes & 0.00144 & 0.8787 & \(<0.001\)\\
60 s & AI & 0.8211 & 0.8314 & \(-0.0102\) & No  & 0.00205 & 0.8798 & \(<0.001\)\\
\bottomrule
\end{tabular}
\begin{minipage}{0.94\linewidth}\footnotesize
Drop is full-sample \(R^2\) minus mean \(R^2\) after permuting the regime
representation within the training procedure.  ``Survives'' indicates a
positive permutation drop.  MZ reports the Mincer-Zarnowitz intercept \(a\),
slope \(b\), and joint calibration-test \(p\)-value.
\end{minipage}
\end{table}

\subsection{Leading-Signal and Causal Falsification}
\label{sec:results:causal}

We finally test whether covariance-state motion anticipates volatility or
merely tracks it.  The event study initially suggests a lead: dial velocity
peaks two days before the 50 largest VSTOXX events, realised volatility peaks
on the event day, and the dial level and VSTOXX peak one day later.
Acceleration does not share the same event-study timing, although its
weak negative lead-lag correlation reaches its extremum one day before
realised volatility and is classified as coincident in the corrected output.
These patterns are therefore suggestive but not sufficient evidence of predictive
precedence.
Derivatives of a persistent noisy state can peak before its level by
construction, so temporal ordering alone is not evidence of causality.
Event alignment can also favour episodes with similar pre-event paths,
strengthening the need for chronological out-of-sample arbitration.

These final tests assess whether the covariance state genuinely leads
volatility or simply moves with it.  In the canonical 30-second AI cell,
velocity and acceleration predict realised volatility
(\(p=3.87\times10^{-4}\) and \(p=0.0233\)), but realised volatility also
predicts both measures (\(p=1.40\times10^{-4}\) and
\(p=1.22\times10^{-13}\)).  The relationship is therefore bidirectional and
does not establish causality.  Chronological out-of-sample results provide a
clearer conclusion.  Across all six duration-metric cells, adding dial
velocity and acceleration jointly to daily persistence improves incremental
\(R^2\) by no more than 0.0071.  Velocity alone clears both block-bootstrap
intervals in the three LE cells but in none of the AI cells, while acceleration
gains are small and inconsistent.  The isolated LE result therefore does not
satisfy the predeclared requirement of agreement across geometries.  In the
canonical cell, velocity and acceleration jointly add 0.00269
\([{-}0.00824,0.01181]\), while including the dial level raises this to
0.00366 \([{-}0.00650,0.01323]\).  These are the primary five-day
moving-block intervals; the corresponding ten-day intervals,
\([{-}0.00785,0.01207]\) and \([{-}0.00762,0.01337]\), lead to the same
conclusion.  Every combined-derivative interval across the six cells includes
zero.  The estimates use 732 out-of-sample days after fold-boundary purging
and a one-day embargo, with 2,000 bootstrap replications.  Critical-slowing-down screening reaches the same broad
conclusion: it identifies two of four crises but also flags 33.3\% of 39
random dates (\(p=0.407\)), with \(p\)-values ranging from 0.36 to 0.77 across
cells.  Overall, the covariance dial has a genuine intraday forecasting
advantage, but the evidence does not support interpreting it as a causal or
crisis-warning indicator at the daily horizon.

\section{Conclusion}\label{sec:conclusion}

Markets may move continuously even when the models used to describe them do
not. For the deep limit order book examined here, conventional market regimes
appear less like naturally separated states and more like convenient
partitions of a continuously evolving stress process. This distinction matters: imposing labels too early
can turn gradual movement into artificial transitions, conceal meaningful
variation within categories, and produce boundaries that appear convincing in
sample but fail as market conditions change.

The hierarchical results show that broad bands can usefully summarise market
stress, but finer classifications increasingly distort the underlying
continuum, replacing meaningful distance and direction with unstable
boundaries that improve in-sample separation while weakening out-of-sample
prediction. This explains the apparent success and eventual failure of
discrete classifications.

The continuous representation preserves ordering
without relying on unstable boundaries. Predictive information shifts
smoothly between slow and fast components as the decomposition timescale
changes, with no privileged boundary between macro and micro dynamics, while
the residual analysis shows that the state is not strictly one-dimensional.
The micro-subspace adds significant predictive value, with the largely linear
forecasting relationship indicating that this information originates from the
covariance state, not model complexity.
Discrete micro-regimes do not recover this information,
confirming that the loss cannot be repaired merely by adding more hierarchical
labels. Even a carefully constructed hierarchy therefore cannot fully
represent a state space whose predictive structure changes smoothly.

Our research provides a practical approach showing how continuous and discrete
representations can complement each other in regime detection and where each
is most appropriate. The relevant transition is not a universal time cutoff,
but the point at which discretisation improves operational clarity without
sacrificing stability or predictive information. Rather than choosing a regime
count in advance, research should focus on comparing geometric evidence,
boundary stability, chronological transfer, and predictive loss across
observation and forecast horizons. A discrete representation is justified where broad partitions
remain stable and transfer out of sample; a continuous representation is
preferable where predictive information survives in the coordinate but
disappears after classification. The transition from a continuous to a
discrete representation must be identified empirically across timescales, not
imposed as a modelling assumption. This perspective has direct
applications beyond regime classification. In simulation, continuous state
dynamics can generate gradual stress accumulation, mean reversion, and
realistic movement between market conditions without artificial jumps between
labels. In risk management, the dial can support live monitoring of market
stress while discrete thresholds remain available for escalation and
communication. On trading desks, it could condition execution, hedging,
exposure, or liquidity decisions on both the current stress level and the
rate at which conditions are changing. Tracking a continuous path may reduce
the delayed reactions and unstable signal switching produced when observations
repeatedly cross rigid regime boundaries. These applications require
dedicated economic evaluation and should not be interpreted as evidence of
trading profitability from the present study.

Future research will develop continuous-regime detection methods that
distinguish ordinary movement, temporary displacement, and structural change,
and test whether the observed structure generalises across markets and crisis
periods. This will help establish when the continuum persists and when
genuinely discrete behaviour emerges.

This research contributes to a broader shift in how market regimes are
understood.
Once the focus shifts from assigning labels to measuring how market conditions
evolve and how their representation transitions from continuous to discrete,
the whole vocabulary of regimes changes. The central question therefore
changes from ``Which state is the market in?'' to ``Where is the market on the
continuum, how quickly is it moving, how persistent is that movement, and when
does discretising it become useful or destructive?'' This change in
perspective preserves more of the information contained in market data. By
separating the geometry of the market state from the labels imposed on it,
this research provides a foundation for developing more flexible regime models
for forecasting, simulation, risk management, and live market monitoring.

\section*{Data and code availability}
The raw FESX limit-order-book records are supplied under an exchange licence
and cannot be redistributed by the authors.  Access therefore requires an
independent licence from the Deutsche B\"orse.
A structured GitHub repository containing the non-proprietary tables, figures,
dashboards, result summaries, and validation materials supporting this study
has been created at
\externallink{https://github.com/anjalitwt009/Latent-Continuum-of-Regimes-in-Limit-Order-Book-Dynamics}{GitHub repository}.
Licensed raw FESX records and any derived files whose redistribution would
violate the Deutsche B\"orse licence are excluded.

\bibliographystyle{rQUF}
\bibliography{paper}

\clearpage
\appendix
\section{Covariance-State Foundation and Metric Robustness}\label{app:canonical}

\subsection{Native-grid rule and mask precedence}

\begin{table}[H]
\centering
\caption{Canonical analysis choices and their paired checks.}
\label{tab:app_canonical}
\small
\begin{tabular}{lll}
\toprule
Object & Canonical choice & Paired or qualifying check\\
\midrule
Windowing & Native 30/45/60 s & Cross-duration agreement\\
Covariance & OAS, $15\times15$ & SPD and conditioning checks\\
Sample & Native union clean mask & Unmasked foundation\\
Geometry & AI headline & LE parallel evaluation framework\\
Representation & Continuous oriented axis & 4/8/16/deep bands\\
Evaluation & Chronological OOS & Block-bootstrap intervals\\
Causal arbiter & OOS increment & Event study and Granger diagnostics\\
\bottomrule
\end{tabular}
\end{table}

\subsection{Log-Euclidean and affine-invariant geometry}

\begin{table}[H]
\centering
\caption{Comparison of Log-Euclidean and affine-invariant covariance geometry.}
\label{tab:app_geometry_comparison}
\scriptsize
\begin{tabular}{@{}p{0.18\linewidth}p{0.35\linewidth}p{0.39\linewidth}@{}}
\toprule
Property & Log-Euclidean (LE) & Affine-invariant (AI)\\
\midrule
Distance
& \(\|\log\Sigma_t-\log\Sigma_u\|_F\);
equation~\eqref{eq:method_dle}
& \(\|\log(\Sigma_t^{-1/2}\Sigma_u\Sigma_t^{-1/2})\|_F
=\{\sum_j\log^2\gamma_{j,t,u}\}^{1/2}\);
equation~\eqref{eq:method_dai}\\

Mean
& \(\bar\Sigma_{\mathrm{LE}}
=\exp\{n^{-1}\sum_i\log\Sigma_i\}\); closed form
& Karcher/Fr\'echet mean
\(\arg\min_M\sum_i d_{\mathrm{AI}}^2(M,\Sigma_i)\); iterative;
equations~\eqref{eq:method_karcher_objective}-\eqref{eq:method_karcher_update}\\

Tangent coordinate
& \(\phi(\Sigma)=\operatorname{vech}_{\sqrt2}(\log\Sigma)
\in\mathbb R^{120}\), based at the identity
& \(Z_t^{(\mathrm{AI})}
=\operatorname{vech}_{\sqrt2}
\{\log(M^{-1/2}\Sigma_tM^{-1/2})\}\in\mathbb R^{120}\),
based at Karcher mean \(M\)\\

Scale/unit sensitivity
& Sensitive to rescaling and linear recombination of the input variables
& Invariant to nonsingular rescaling and linear recombination\\

Congruence invariance
& Not generally congruence invariant
& Congruence invariant; means and centroids are equivariant\\

Numerical cost
& Lower: one matrix logarithm per state followed by Euclidean operations
& Higher: generalised eigen decompositions for distances and iterative Karcher
updates, performed in double precision for metric comparisons\\

Graded perturbation
& None
& \(10^{-6}\) to \(3\times10^{-6}\) times the mean diagonal for the 45 and
60-second paths only; none at 30 seconds\\

Role
& Exact-isometry operational tangent representation and robustness benchmark
& Primary metric for headline, metric-sensitive analysis\\

Where used
& Operational continuum coordinate, tangent-space calculations, PCA, and
Log-Euclidean hierarchy and robustness checks
& Primary distances, means, certificates, and hierarchy; separately estimated
Karcher-centred principal-geodesic coordinate \(s_t^{(\mathrm{AI})}\) where stated\\
\bottomrule
\end{tabular}
\end{table}

\subsection{Feature inventory}

\begin{table}[H]
\centering
\caption{Fixed-order event-level variables used to construct each covariance state.}
\label{tab:app_covariance_feature_inventory}
\small
\begin{tabular}{@{}r p{0.22\linewidth} p{0.45\linewidth} p{0.18\linewidth}@{}}
\toprule
\# & Variable & Description & Group\\
\midrule
1  & \texttt{bid\_price\_1}       & Best bid price (L1) & Price\\
2  & \texttt{ask\_price\_1}       & Best ask price (L1) & Price\\
3  & \texttt{mid}                 & Mid-price \(\tfrac{1}{2}(b_1+a_1)\) & Price\\
4  & \texttt{spread}              & Quoted spread \(a_1-b_1\) & Cost\\
5  & \texttt{relative\_spread}    & Spread divided by mid-price & Cost\\
6  & \texttt{microprice}          & Size-weighted microprice & Price/pressure\\
7  & \texttt{bid\_vol\_1}         & Displayed L1 bid quantity & Depth\\
8  & \texttt{ask\_vol\_1}         & Displayed L1 ask quantity & Depth\\
9  & \texttt{trade\_volume}       & Traded volume associated with the observation & Activity\\
10 & \texttt{depth\_imbalance\_5} & Five-level depth imbalance & Depth/imbalance\\
11 & \texttt{imbalance\_1}        & Level-1 imbalance & Imbalance\\
12 & \texttt{imbalance\_2}        & Level-2 imbalance & Imbalance\\
13 & \texttt{imbalance\_3}        & Level-3 imbalance & Imbalance\\
14 & \texttt{ofi\_weighted}       & Depth-weighted five-level order-flow imbalance
                                      using depth decay & Order flow\\
15 & \texttt{price\_pressure}     & Clipped price pressure & Order flow\\
\bottomrule
\end{tabular}
\end{table}

\subsection{Window-store feature families}

\begin{table}[H]
\centering
\caption{Composition of the 143-variable window store by feature family.}
\label{tab:app_window_feature_families}
\small
\begin{tabular}{@{}p{0.34\linewidth} r p{0.50\linewidth}@{}}
\toprule
Family & \# & Examples\\
\midrule
Order flow/imbalance
& 29 & \texttt{ofi\_window}, \texttt{imbal\_1-10},
\texttt{price\_pressure\_mean}, \texttt{vpin\_proxy}\\
Rolling robust z-scores/warm-up
& 17 & \texttt{spread\_z}, \texttt{vol\_z}, \texttt{ofi\_z},
\texttt{feat\_warmup}\\
Lags
& 16 & \texttt{mid\_return\_lag1/2},
\texttt{realized\_vol\_lag1/2}\\
Price/return
& 14 & \texttt{mid\_open/close/high/low}, \texttt{mid\_return},
\texttt{vwap}\\
Spread/cost
& 13 & \texttt{spread\_median/max/min},
\texttt{effective\_spread\_mean}, \texttt{roll\_spread}\\
Depth/book shape
& 13 & \texttt{depth\_bid/ask\_\{3,5,10\}},
\texttt{book\_slope\_ratio\_mean}, \texttt{depth\_gini}\\
Realised variation/volatility
& 12 & \texttt{realized\_var}, \texttt{bipower\_var},
\texttt{parkinson\_vol}, \texttt{garman\_klass\_vol}\\
Trading activity
& 12 & \texttt{n\_upticks}, \texttt{n\_trades},
\texttt{events\_per\_sec}, \texttt{arrival\_cv}\\
Calendar position
& 6 & \texttt{minutes\_since\_open}, \texttt{is\_first\_30min},
\texttt{day\_of\_week}\\
Liquidity/impact
& 4 & \texttt{kyle\_lambda}, \texttt{amihud\_illiq},
\texttt{message\_to\_trade\_ratio}\\
VSTOXX/external
& 4 & \texttt{vstoxx\_level}, \texttt{vstoxx\_return},
\texttt{fesx\_vstoxx\_corr}\\
Jump measures
& 3 & \texttt{jump\_component}, \texttt{jump\_flag},
\texttt{jump\_ratio}\\
\midrule
Total & 143 & \\
\bottomrule
\end{tabular}
\begin{minipage}{0.94\linewidth}\footnotesize
Counts are mutually exclusive and sum to 143.
These window-level variables support anomaly detection, forecasting targets, and validation;
they do not enter \(\Sigma_t\) directly. Table~\ref{tab:app_covariance_feature_inventory}
separately identifies the 15 raw event-level covariance inputs, including any similarly named
analogues.
\end{minipage}
\end{table}

\section{Anomaly-Gate Validation and Clean-Sample Robustness}

\subsection{Isolation Forest anomaly-feature inventory}
\label{app:iforest_features}\linkdest{app:iforest_features}

\begin{table}[H]
\centering
\caption{Isolation Forest anomaly-feature inventory.}
\label{tab:app_iforest_feature_inventory}
\scriptsize
\setlength{\tabcolsep}{3pt}
\renewcommand{\arraystretch}{1.08}
\begin{tabular}{@{}p{0.25\linewidth} r p{0.64\linewidth}@{}}
\toprule
Feature family & \# & Features\\
\midrule
Volatility and market efficiency
& 7 & \texttt{realized\_vol}, \texttt{realized\_vol\_z},
\texttt{variance\_ratio}, \texttt{jump\_component},
\texttt{return\_kurt}, \texttt{return\_skew},
\texttt{return\_autocorr}\\
Spread and liquidity
& 5 & \texttt{spread\_std}, \texttt{spread\_wide\_frac},
\texttt{spread\_z}, \texttt{effective\_spread\_mean},
\texttt{realised\_vs\_quoted}\\
Price impact and microstructure
& 4 & \texttt{kyle\_lambda\_z}, \texttt{amihud\_z},
\texttt{vwap\_dev\_z}, \texttt{roll\_spread}\\
Order flow and toxicity
& 5 & \texttt{ofi\_toxicity}, \texttt{vpin\_proxy},
\texttt{trade\_imbalance\_z}, \texttt{ofi\_level\_corr},
\texttt{ofi\_weighted\_z}\\
Book depth and shape
& 5 & \texttt{depth\_imbalance\_5}, \texttt{imbal\_entropy},
\texttt{depth\_gini}, \texttt{depth\_concentration},
\texttt{depth\_recovery\_mean}\\
Activity and arrivals
& 5 & \texttt{arrival\_cv}, \texttt{arrival\_cv\_z},
\texttt{activity\_z}, \texttt{message\_to\_trade\_ratio},
\texttt{null\_trade\_fraction}\\
Trade momentum and informed flow
& 3 & \texttt{mid\_moved\_frac}, \texttt{price\_pressure\_mean},
\texttt{post\_trade\_spread\_spike\_m}\\
Liquidity resilience
& 2 & \texttt{bid\_queue\_chg\_sum},
\texttt{ask\_queue\_chg\_sum}\\
Ten-level book and jump-aware features
& 4 & \texttt{depth\_imbalance\_10},
\texttt{depth\_imbalance\_10\_z}, \texttt{ofi\_weighted\_10\_z},
\texttt{jump\_ratio}\\
\midrule
Total & 40 & \\
\bottomrule
\end{tabular}
\end{table}

\section{Continuous Stress Dynamics and Online Monitoring}

\subsection{Stress-relaxation formulas}
\label{app:relaxation_formulas}\linkdest{app:relaxation_formulas}

For each native duration, the discrete Ornstein-Uhlenbeck approximation is
estimated from consecutive observations belonging to the same trading day:
\begin{equation}
s_{t+1}=a+bs_t+\varepsilon_{t+1},
\qquad 0<b<1
\label{eq:method_ou}
\end{equation}
Here, \(s_t\) is the continuous stress index, \(a\) is the intercept, \(b\) is
one-window persistence, and \(\varepsilon_{t+1}\) is the next-window
innovation. Writing \(b=e^{-\theta\Delta}\), the mean-reversion rate and
long-run stress level are
\begin{equation}
\theta=-\frac{\log b}{\Delta},
\qquad
\mu=\frac{a}{1-b}
\label{eq:method_ou_parameters}
\end{equation}
where \(\Delta\) is the native 30, 45 or 60-second interval. The implied
half-life is
\begin{equation}
\tau_{1/2}
=\frac{\log2}{\theta}
=-\frac{\Delta\log2}{\log b}
\label{eq:method_ou_halflife}
\end{equation}

For the nonparametric calculation, \(q_t\in\{1,\ldots,12\}\) denotes the
ordered quantile cell containing \(s_t\). Same-day transitions estimate
\begin{equation}
P_{ab}
=
\frac{
\sum_t\mathbf 1\{q_t=a,\ q_{t+1}=b,\ d(t+1)=d(t)\}
}{
\sum_t\mathbf 1\{q_t=a,\ d(t+1)=d(t)\}
},
\qquad
\sum_bP_{ab}=1
\label{eq:app_stress_transition}
\end{equation}
where \(P_{ab}\) is the estimated probability of moving from cell \(a\) to
cell \(b\) over one native interval and \(d(t)\) identifies the trading day.
For each non-unit eigenvalue \(\lambda_j\) of \(P\), the implied relaxation
time is
\begin{equation}
\tau_j=-\frac{\Delta}{\log|\lambda_j|}
\label{eq:method_transfer_time}
\end{equation}
The leading non-unit mode is compared with \(\tau_{1/2}\); agreement within a
factor of four constitutes the predeclared convergence check.  The complete
continuous stress-dynamics certificate and the one-window predictive-coincidence
diagnostics are reported in the supplementary materials.

\section{Intraday Volatility Forecasting}

\subsection{Realised-variation targets and forecast horizons}
\label{app:forecast_targets}\linkdest{app:forecast_targets}

\begin{table}[H]
\centering
\caption{Realised-variation quantities and forecast-horizon specification.}
\label{tab:app_forecast_targets}
\small
\begin{tabular}{@{}p{0.24\linewidth} p{0.47\linewidth} p{0.22\linewidth}@{}}
\toprule
Quantity & Definition & Role\\
\midrule
Realised variance and volatility
& \(\mathrm{RV}_t=\sum_i r_i^2\) and
\(\mathrm{RVol}_t=\sqrt{\mathrm{RV}_t}\), using within-session log-mid-price
returns with the first session return set to zero
& Forecast target and persistence benchmark\\

Bipower variation
& \(\mathrm{BV}_t=(\pi/2)\sum_i |r_i||r_{i-1}|\)
& Jump-robust continuous-variation estimate\\

Jump component
& \(J_t=\max\{\mathrm{RV}_t-\mathrm{BV}_t,0\}\)
& Non-negative discontinuity measure\\

Forward horizons
& Next complete native window; 30, 45, and 60 seconds; 2, 5, 10, 15, and
30 minutes; and 1 hour
& Multi-horizon forecasting\\

Constraints
& Targets remain within the same trading session, and no forecast horizon is
finer than the native covariance duration
& Temporal-alignment and leakage control\\
\bottomrule
\end{tabular}
\begin{minipage}{0.94\linewidth}\footnotesize
These quantities are constructed separately from the 15 event-level inputs to
\(\Sigma_t\). Realised variance enters the jump decomposition, while realised
volatility supplies the volatility-scale forecasting outcome.
\end{minipage}
\end{table}

\end{document}